\documentclass[prd,nofootinbib,superscriptaddress]{revtex4-2}
\usepackage[a4paper,left=1.5cm,right=1.5cm,top=3cm,bottom=3cm]{geometry}

\usepackage{tensor}
\usepackage{xcolor}
\usepackage{enumerate}
\usepackage{framed}
\usepackage{mdframed}
\usepackage{amsmath}
\usepackage{amsfonts}
\usepackage{mathrsfs}
\usepackage{amssymb}
\usepackage{graphicx, rotating}
\usepackage{epsfig}
\usepackage{latexsym}
\usepackage{graphicx}
\usepackage{color}
\usepackage{amsmath,bm,amssymb}
\usepackage{slashed}
\usepackage{hyperref}
\usepackage[font=footnotesize,labelfont=]{caption}
\hypersetup{colorlinks=true, citecolor=bluscuro, linkcolor=black, urlcolor=bluscuro}
\definecolor{rossos}{cmyk}{0,1,1,0.55}
\definecolor{bluscuro}{rgb}{0.15, 0.2, .85}
\definecolor{bluchiaro}{cmyk}{1,.3,0.,0.1}
\usepackage[T1]{fontenc} 
\usepackage{subfig}

\usepackage{tabularx}

\usepackage{tcolorbox}

\def\0{\vec{0}}

\newcommand{\co}{ \text{\tiny cut-off} }
\newcommand{\bea}{\begin{eqnarray}}
	\newcommand{\eea}{\end{eqnarray}}
\def\beq{\begin{equation}}
	\def\eeq{\end{equation}}

\def\d{{\rm d}}

\def\beqa{\begin{eqnarray}}

	\def\eeqa{\end{eqnarray}}
\def\lsim{\mathrel{\rlap{\lower4pt\hbox{\hskip0.5pt$\sim$}}
		\raisE_1pt\hbox{$<$}}}         
\def\gsim{\mathrel{\rlap{\lower4pt\hbox{\hskip0.5pt$\sim$}}
		\raisE_1pt\hbox{$>$}}}         

\def\d{{\rm d}}

\def\d{{\rm d}}

\def\PBH{\text{\tiny PBH}}
\def\ABH{\text{\tiny ABH}}
\def\DM{\text{\tiny DM}}

\def\eeqa{\end{eqnarray}}
\numberwithin{equation}{section}

\def\bq{\begin{quote}}
\def\eq{\end{quote}}

 at 10truept
\newcommand{\arXiv}[2]{\href{http://arxiv.org/pdf/#1}{{\tt [#2/#1]}}}
\newcommand{\arXivold}[1]{\href{http://arxiv.org/pdf/#1}{{\tt [#1]}}}

\newcommand{\llp}{\left [}
\newcommand{\rrp}{\right ]}

\newcommand{\lp}{\left (}
\newcommand{\rp}{\right )}

\newcommand{\be}{\begin{equation}\begin{aligned}}
	\newcommand{\ee}{\end{aligned}\end{equation}}

\usepackage[normalem]{ulem}

\def\ii{{\text{\tiny i}}}

\def\cs2{c_{\rm{s}}^2}

\def\U0{{\bar U_0}}
\def\bi{\begin{itemize}}
\def\ei{\end{itemize}}

\def\U{{\cal{U}}}

\begin{document}

\title{
Bayesian Evidence for Both Astrophysical and Primordial Black Holes:\\
Mapping the GWTC-2 Catalog to Third-Generation Detectors
}

\author{V. De Luca}
\address{D\'epartement de Physique Th\'eorique and Centre for Astroparticle Physics (CAP), Universit\'e de Gen\`eve, 24 quai E. Ansermet, CH-1211 Geneva, Switzerland}

\author{G. Franciolini}
\address{D\'epartement de Physique Th\'eorique and Centre for Astroparticle Physics (CAP), Universit\'e de Gen\`eve, 24 quai E. Ansermet, CH-1211 Geneva, Switzerland}

\author{P. Pani}
\address{Dipartimento di Fisica, 
Sapienza Universit\`a di Roma, Piazzale Aldo Moro 5, 00185, Roma, Italy}
\address{INFN, Sezione di Roma, Piazzale Aldo Moro 2, 00185, Roma, Italy}

\author{A.~Riotto}
\address{D\'epartement de Physique Th\'eorique and Centre for Astroparticle Physics (CAP), Universit\'e de Gen\`eve, 24 quai E. Ansermet, CH-1211 Geneva, Switzerland}

\date{\today}

\begin{abstract}
\noindent
We perform a hierarchical Bayesian analysis of the GWTC-2 catalog to investigate the mixed scenario in which the merger events are explained by black holes of both astrophysical and primordial origin. 
For the astrophysical scenario we adopt the phenomenological model used by the LIGO/Virgo collaboration and we include  the correlation between different parameters inferred from data, the role of the spins in both the primordial and astrophysical  scenarios, and the impact of accretion in the primordial scenario.  
Our best-fit mixed model has a strong statistical evidence relative to the single-population astrophysical model, thus 
supporting  the coexistence of  populations of black-hole mergers of two different origins. 
In particular, our results indicate that the astrophysical mergers account for roughly four  times the number of primordial black hole events and predict that third-generation detectors,  such as the Einstein Telescope and Cosmic Explorer,  should detect up to hundreds of mergers from primordial black hole binaries at redshift $z\gtrsim30$.

\end{abstract}

\maketitle

\section{Introduction}
\renewcommand{\theequation}{1.\arabic{equation}}
\setcounter{equation}{0}
\label{intro}

\noindent
The several detections of the Gravitational Waves~(GWs) coming from black hole~(BH) mergers 
during the first three observing runs performed by the LIGO/Virgo Collaboration~(LVC)~\cite{LIGOScientific:2018mvr,Abbott:2020niy}
has ignited the interest in understanding the origin and the properties of the Binary Black Hole~(BBH) population~\cite{Barack:2018yly}.

Multiple formation scenarios have been proposed to explain the formation of the observed BBHs and their merger in the local universe. They can be broadly distinguished in two categories.
One is the Astrophysical Black Hole~(ABH) formation scenario (recently reviewed in Refs.~\cite{Mandel:2018hfr,Mapelli:2018uds}) where the BHs form at the end of the stellar evolution in the late-time universe, forming binaries either in isolation through a common envelope phase or dynamically through three-body encounters in dense stellar clusters.
The other possibility is provided by the Primordial Black Hole~(PBH) scenario, firstly proposed in the late '70s in Refs.~\cite{1967SvA....10..602Z,Hawking:1974rv,Chapline:1975ojl}, where the compact objects are formed in the radiation dominated early-universe out of the collapse of large overdensities~\cite{s00,s1}, and binaries are assembled via random gravitational decoupling from the Hubble flow before the matter-radiation equality (see Refs.~\cite{Sasaki:2018dmp,Green:2020jor} for some recent reviews).
Additionally, PBHs in different mass ranges could contribute to a sizeable fraction $f_\PBH\equiv \Omega_\PBH/\Omega_\DM$ of the Dark Matter~(DM)~\cite{Carr:2020gox}. This has 
motivated recent efforts in confronting the PBH scenario with the GW data~\cite{Sasaki:2016jop,Bird:2016dcv,Clesse:2016vqa, Wang:2016ana,
Ali-Haimoud:2017rtz, raidal, raidalsm,ver, Gow:2019pok,  Fernandez:2019kyb,Hall:2020daa,Wong:2020yig,DeLuca:2020jug,Hutsi:2020sol} in search for evidences for  PBHs. Also, PBHs may be responsible for the formation of binaries like GW190521~\cite{massgapevent} where at least one component confidently falls in the so-called upper mass gap~\cite{DeLuca:2020sae,Kritos:2020wcl} in which ABH binaries are not generally expected to be produced.

One recent effort of the community has been on inferring the properties of PBHs, e.g. their mass distribution and abundance, assuming that all BBH events detected so far have a primordial origin. While this possibility is certainly still allowed by current constraints (see, e.g., Ref.~\cite{Wong:2020yig} for the most updated analysis using the second GW transient catalog (GWTC-2)), it is plausible that all the events may not be explained by means of a single population, either astrophysical~\cite{Zevin:2020gbd} or primordial~\cite{Hutsi:2020sol}.  
In order to distinguish the contribution of populations of different origin in the data, one needs to rely on different characteristic features which may be a smoking gun for either PBHs or ABHs. 
One promising feature of the PBH model is the peculiar evolution of the merger rate, which is monotonically increasing with redshift. This differs from what is expected from an ABH population, whose merger rate should peak around redshift of a few. 
This prediction can be employed by future third-generation~(3G) detectors like the Einstein Telescope~(ET)~\cite{et} and Cosmic Explorer~(CE)~\cite{ce} to confidently distinguish PBHs from other astrophysical formation channels, in particular from BHs formed from the first stars~\cite{Schneider:1999us,Schneider:2001bu,Schneider2003}, which might merge also at high redshift~\cite{Ng:2020qpk,Liu:2020ufc,Valiante2021}.

As a first step towards this future direction, in this paper we aim at inferring the binary population properties 
considering both the PBH and ABH formation channels. We restrict our description of the astrophysical sector to the one 
used by the LVC by means of a phenomenological model and mix it with the PBH model by performing a hierarchical Bayesian 
analysis on the GWTC-2 catalog.  

The purpose of this work is twofold. First, we wish to quantify what 
fraction of the current observed events may be ascribed to PBHs. Secondly, we wish to use this mixing fraction to make a 
forecast for the expected number of events to be observed by 3G detectors at high redshifts, as well as the 
corresponding redshift dependence of the merger rate in this mixed ABH-PBH formation scenario. 

Our hierarchical Bayesian analysis improves over the recent analysis of Ref.~\cite{Hutsi:2020sol} by including the correlation between different parameters inferred from data, the role of the spins in both the PBH and ABH scenarios, and the impact of accretion in the PBH model.

The paper is organised as follows. In Sec.~\ref{sec1} we summarise the hierarchical Bayesian framework adopted in this paper to perform the inference on the GWTC-2 catalog. In Secs.~\ref{sec2} and~\ref{sec3} we describe the PBH and ABH models considered in this paper, while also reporting both single-population inferences. 
Sec.~\ref{sec4} contains our main result showing the mixed population inference and Sec.~\ref{sec5} draws the forecasted detectability of the uncontaminated PBH subpopulation at high redshift by 3G detectors. We conclude in Sec.~\ref{sec:conclusions} with some future prospects, whereas the Appendices are devoted to some technical aspects of both the binary detectability and the PBH accretion model, respectively.

\section{Bayesian inference}\label{sec1}
\renewcommand{\theequation}{2.\arabic{equation}}
\setcounter{equation}{0}

\noindent
In this work we perform a hierarchical Bayesian inference analysis to select the best fitting models given the GWTC-2 dataset in the full PBH, full ABH and mixed ABH-PBH binary populations scenarios.
In doing so, we  closely follow the procedure used in~\cite{Hall:2020daa,Wong:2020yig} and references therein.
In this section we provide a short summary of the hierarchical Bayesian inference framework, highlighting the key inputs coming from both the models considered and the dataset. The knowledgeable reader may jump directly to the following sections where the models considered in this work are presented.

On general grounds, a hierarchical Bayesian analysis returns the posterior distribution of a series of model hyperparameters ${\bm \lambda}$ based on the observed dataset $\bm{d}$. In the following, we will also define the intrinsic event parameters, such as masses and redshift of the binary mergers, as $\bm{\theta}$.
In practice, the posterior distribution can be inferred from the data by sampling over the hyperparameter space $\bm{\lambda}$ the function
\begin{align}
p(\bm{\lambda}|\bm{d}) \propto \pi(\bm{\lambda})\int p(\bm{d}|\bm{\theta})
p_\text{\tiny pop}(\bm{\theta}|\bm{\lambda})\d \bm{\theta},
\label{eq:HBA}
\end{align}
where $p(\bm{d}|\bm{\theta})$ is the single-event likelihood, $\pi(\bm{\lambda})$ is the prior on the hyperparameters, and $p_\text{\tiny pop}(\bm{\theta}|\bm{\lambda})$ is the \textit{population likelihood}, equivalent to a prior on the intrisic parameters $\bm{\theta}$ but parametrized by the model hyperparameters $\bm{\lambda}$. 
The parameters describing single events (e.g. masses, redshifts) are also referred to as {\it event parameters}, while the hyperparameters describing the entire sample (e.g. the fraction of DM in PBHs) can be also referred to as {\it population parameters}.  For clarity, in Table~\ref{Tb:parameters} we provide a summary of all the parameters defined in this paper.

In order to speed up the computation, such an integration in Eq.~\eqref{eq:HBA} is written as a weighted average over the posterior samples.
The likelihood function is then computed according to the following equation 
 (see for example  Ref.~\cite{selectioneffects}  for its derivation)
\begin{align}
p(\bm{\lambda}|\bm{d}) =
 \pi(\bm{\lambda})e^{- N_{\rm det} ({\bm \lambda})} 
 \llp N({\bm \lambda})\rrp^{N_{\rm obs}}
 \prod_{i=1}^{N_{\rm obs}}\frac{1}{{\cal S}_i}\sum_{j=1}^{{\cal S}_i} \frac{p_\text{\tiny pop}(^j\bm{\theta}_i|\bm{\lambda})}{\pi(^j\bm{\theta}_i)}\,,
\label{eq:populationPosterior_discrete}
\end{align}
where $N_\text{\tiny obs}$ is the number of events in the catalog, i.e. the number of detections, $N({\bm \lambda})$ is the number of events in the model characterised by the set of hyperparameters ${\bm \lambda}$, $N_\text{\tiny det} ({\bm \lambda})$ is the expected number of {\it observable} events in the model characterised by the  hyperparameters ${\bm \lambda}$ computed by accounting for the experimental selection bias \cite{selectioneffects}, and ${\cal S}_i$ is a normalisation factor which depends on the length of the posterior dataset of each event in the catalog.
Finally, $\pi(\bm{\theta})$ is the prior on the binary parameters used by the LVC when performing the parameter estimation. This prior is removed to extract the values of the single-event likelihood ensuring only the informative part of the event posterior is used.

\begin{table}
\renewcommand{\arraystretch}{1.3}
\caption{Relevant parameters in the analysis.}
\begin{tabularx}{\columnwidth}{l X}
\hline
\hline
Event parameters $\bm{\theta}$ &  \\
$m_1$ & Source-frame primary mass \\
$m_2$ & Source-frame secondary mass \\
$\chi_{\rm eff}$ & Effective spin  \\
$z$ & Merger redshift \\
\hline
Hyperparameters $\bm{\lambda}$ of the PBH model &  \\
$M_c$ & Peak reference mass of the lognormal distribution\\
$\sigma$ & Variance of the lognormal mass distribution\\
$f_\PBH$ & Fraction of PBHs in DM at formation \\
$z_\co$ & Accretion cut-off redshift \\
\hline
Hyperparameters $\bm{\lambda}$ of the ABH model &  \\
$R_0$ & Integrated merger rate at $z=0$ \\
$\kappa$ & Merger rate evolution in redshift, defined as $R(z) \propto (1+z)^\kappa$\\
$\alpha$ & Exponent of the total mass factor in the differential rate \\
$\beta$ & Exponent of the symmetric mass ratio factor\\
$\zeta$ & ABH mass function power law scaling $\psi (m) \propto m^{-\zeta}$\\
$m_\text{\tiny min}$ & Minimum ABH mass (step-like cut-off) \\
$m_\text{\tiny max}$ & Maximum ABH mass (step-like cut-off) \\
$\mu_{\chi_\text{\tiny eff}}$ & Mean value of $\chi_\text{\tiny eff}$ in the ``Gaussian''  spin model\\
$\sigma_{\chi_\text{\tiny eff}}$ & Width of $\chi_\text{\tiny eff}$ in the ``Gaussian''  spin model\\
\hline
Inferred parameters in the mixed model \hspace{.6 cm} &  \\
$f_\text{\tiny P} \equiv N^\text{\tiny det}_\text{\tiny PBH}/(N^\text{\tiny det}_\text{\tiny ABH}+ N^\text{\tiny det}_\PBH)$ &  Fraction of PBH binaries  \\
$f_\text{\tiny A} \equiv N^\text{\tiny det}_\text{\tiny ABH}/(N^\text{\tiny det}_\text{\tiny ABH}+ N^\text{\tiny det}_\PBH) \equiv 1- f_\text{\tiny P}$ &  Fraction of ABH binaries \\
\hline
\hline
\end{tabularx}
\label{Tb:parameters}
\end{table}

We remark that the last sum in Eq.~\eqref{eq:populationPosterior_discrete} gives the average of the binary parameter distribution over posterior samples. 
The factors proportional to 
\begin{equation}
	p(\bm{\lambda}|\bm{d}) \propto e^{- N_{\rm det} ({\bm \lambda})} \llp N({\bm \lambda})\rrp ^{N_{\rm obs}}
\end{equation}
characterize an inhomogeneous Poisson process~\cite{2019PASA...36...10T,Loredo:2004nn,Taylor:2018iat,Mandel:2018mve} and are responsible for introducing the rate information in the inference.

In our analysis we make the following assumptions for the relevant quantities:
\begin{itemize}
\item the prior $\pi ({\bm \lambda})$ on the hyperparameters ${\bm \lambda}$ of the model is considered flat;
\item for a given model ${\cal M}$ with hyperparameter ${\bm \lambda}$, $p_\text{\tiny pop}(\bm{\theta}|\bm{\lambda})$ is the distribution of the BBH parameters, taken to be $\theta_i =\{ m_1,m_2,z,\chi_\text{\tiny eff}\}$, where $m_i$ is the source-frame  mass of the $i$-th binary component, $z$ is the merger redshift, and 
\begin{equation}
\label{chieff}
\chi_\text{\tiny eff} \equiv \frac{\chi_1 \cos{\alpha_1} + q \chi_2 \cos{\alpha_2}}{1+q}
\end{equation}
is the effective spin parameter, which is a function of the mass ratio $q\equiv m_2/m_1$, of both BH spin magnitudes $\chi_j$ ($j=1,2$, with $0\leq \chi_j\leq 1$), and of their orientation with respect to the orbital angular momentum parametrised by the angles $\alpha_j$. We neglect the precession spin component $\chi_\text{\tiny p}$ in the inference, since this parameter is still poorly reconstructed for most of the GW events~\cite{Abbott:2020gyp};
\end{itemize}

The binary parameter distributions in a given model (either primordial or astrophysical), can be computed from the differential merger rate $\d R/ \d m_1 \d m_2 \d z$ as 
\begin{equation}
p_\text{\tiny pop} (\bm{\theta}|\bm{\lambda}) \equiv \frac{1}{N({\bm \lambda})} \llp  T_\text{\tiny obs} \frac{1}{1+z} \frac{\d V}{\d z} \frac{\d R}{\d m_1 \d m_2} (\bm{\theta}|\bm{\lambda})\rrp,
\label{ppop}
\end{equation}
in terms of the observation time $T_\text{\tiny obs}$, 
while the number of expected detections are defined as the integral of the differential  merger rate as
\begin{equation}
N_\text{\tiny det}(\bm{\lambda})
 \equiv 
  T_\text{\tiny obs}
   \int \d m_1 \d m_2 \d z\, p_\text{\tiny det} (m_1, m_2, z)  \frac{1}{1+z} \frac{\d V}{\d z} \
  \frac{\d R}{\d m_1 \d m_2}  (m_1, m_2, z|\bm{\lambda}) ,
\end{equation}
accounting for the selection bias by introducing the probability of detection $p_\text{\tiny det}$ as defined in Appendix~\ref{app:A}.
As customary, the prefactor $1/(1+z)$ accounts for the clock redshift at the source epoch and $\d V / \d z$ stands for the differential comoving volume factor, see for example~\cite{Dominik:2014yma}.

The analysis is performed by  sampling the likelihood function \eqref{eq:populationPosterior_discrete} in the hyperparameter space by using the Markov chain Monte Carlo software algorithm \texttt{emcee}~\cite{2013PASP..125..306F}.

In the following we will compare how various models are able to explain the GWTC-2 dataset. For a quantitative assessment, we shall use the statistical evidence $Z$. 
Given a model ${\cal M}$, the evidence is defined as the marginal population likelihood computed as the integral of the population posterior $p({\bm \lambda}| {\bm d})$, i.e.
\begin{equation}
Z_{\cal M} \equiv \int \d {\bm \lambda} \, p(\bm{\lambda}|\bm{d}).
\end{equation}
In other words, the evidence is a measure of the support for a given model given the data ${\bm d}$. One can then compare different models by computing the so-called Bayes factors\footnote{
We checked the accuracy of our computation of the Bayes factors by using a nested sampling algorithm, as implemented in DYNESTY  \cite{2020MNRAS.493.3132S}.}
\begin{equation}
{\cal B}^{{\cal M}_1}_{{\cal M}_2} \equiv \frac{ Z_{{\cal M}_1}}{Z_{{\cal M}_2}}.
\end{equation}
According to Jeffreys' scale criterion~\cite{Jeffreys}, a Bayes factor larger than $(10,10^{1.5},10^2)$ would imply a strong, very strong, or decisive evidence in favour of model ${\cal M}_1$ with respect to model ${\cal M}_2$ given the available dataset.

Among all the binary events included in the GWTC-2 catalog, we select the same subset events as used in the LVC population property paper~\cite{Abbott:2020gyp} for a total of $44$ BBH events. Therefore, we discard events with large FAR (false-alarm rate), i.e. GW190426, GW190719, GW190909, and events where the secondary binary component has mass smaller than $3 M_\odot$, i.e. GW170817, GW190425, GW190814, to avoid contamination from putative merger events involving neutron stars. 
Similarly to Ref.~\cite{Wong:2020yig}, we adopt the ``Overall\_posterior'' as provided in~\cite{119} for the GWTC-1 events, and the ``PublicationSamples'' in~\cite{120} for the GWTC-2 events.

Finally, at variance with Ref.~\cite{Hutsi:2020sol}, the present analysis fully accounts for the correlation between the intrinsic event parameters ${\bm \theta}$ in the posterior data and it is not sensitive to the particular choice of single-event priors used by the LVC when performing the parameter estimation.  As shown in Refs.~\cite{pr1,pr2,Huang:2020ysn,pr3} (and in Refs.~\cite{Bhagwat:2020bzh,inprep} in particular for the PBH scenario), the priors may modify the interpretation of some events, also possibly affecting the maximum likelihood analysis as performed in Ref.~\cite{Hutsi:2020sol}.
Furthermore, as far as the PBH model is concerned, we include the effects of accretion onto PBH binaries.
Finally, we also include the spin information in the inference, which is not included in Ref.~\cite{Hutsi:2020sol}.

\section{The primordial black hole model}
\renewcommand{\theequation}{3.\arabic{equation}}
\setcounter{equation}{0}
\label{sec2}

\noindent
For an extensive description of the PBH model we adopt throughout the paper, one can rely on Refs.~\cite{Raidal:2018bbj,DeLuca:2020qqa,Wong:2020yig}. 
We consider the differential merger rate of PBH binaries formed in the early universe 
 as given by~\cite{DeLuca:2020qqa}
\begin{align}
\frac{\d R_\text{\tiny PBH} }{\d m^\ii_1 \d m^\ii_2}
& = 
\frac{1.6 \times 10^6}{{\rm Gpc^3 \, yr}} 
f_\PBH^{\frac{53}{37}} \,
\eta_\ii^{-\frac{34}{37}} 
\lp \frac{t}{t_0} \rp^{-\frac{34}{37}}  
 \lp \frac{M^\ii_\text{\tiny tot}}{M_\odot} \rp^{-\frac{32}{37}}  
S\lp M^\ii_\text{\tiny tot}, f_\PBH,\psi_\ii  \rp
\psi_\ii(m^\ii_1, z_\ii) \psi_\ii (m^\ii_2, z_\ii)
\nonumber \\
&\times 
\exp\llp \frac{12}{37} \int_{t_\ii} ^{t_\text{\tiny cut-off}} \d t \lp \frac{\dot M_\text{\tiny tot}}{M_\text{\tiny tot}} + 2 \frac{\dot \mu}{\mu} \rp
\rrp
\lp\frac{\eta (z_\text{\tiny cut-off})}{\eta (z_\text{\tiny i}) }  \rp^{3/37}
\lp \frac{ M_\text{\tiny tot}(z_\text{\tiny cut-off})}{M_\text{\tiny tot} (z_\text{\tiny i}) } \rp^{9/37} ,
\label{PBHrate}
\end{align}
with $\eta = m_1 m_2/(m_1+m_2)^2$, $M_\text{\tiny tot}= m_1+m_2$, $t_0$ being the current age of the universe, and $S$ being a suppression factor discussed below. In all the expressions the subscript ``i'' indicates quantities at formation epoch, which for a PBH of mass $m$ corresponds to a  typical redshift $z_\ii \simeq 2 \cdot 10^{11} (m/M_\odot)^{-1/2}$.  

The mass function considered in this work is a lognormal mass function at high redshift of the form 
\begin{equation}\label{psi}
\psi_\ii (m| M_c, \sigma)= \frac{1}{m \sigma \sqrt{2 \pi}} \exp \llp - \frac{\ln ^2 (m/M_c)}{2 \sigma ^2}\rrp \,,
\end{equation}
which typically arises when PBHs are formed from the collapse of large overdensities in the radiation dominated early universe with perturbations characterised by a symmetric peak in the comoving curvature power spectrum~\cite{mf1,mf2}. Other possibilities can be considered, see for example Refs.~\cite{Hall:2020daa,Gow:2020cou,Hutsi:2020sol}.

The cut-off redshift $z_\co$ is a hyperparameter of the accretion model which accounts for the uncertainty affecting the strength of accretion close to the reionization epoch, see discussion in Appendix~\ref{app:A}. In this work we follow the approach used in Ref.~\cite{Wong:2020yig} and consider $z_\co$ as a free hyperparameter of the model. For each value of $z_\co$ there exists a one-to-one correspondence between the initial and final masses, $m_\text{\tiny i} \to m$, which can be computed according to the accretion model described in details in Refs.~\cite{DeLuca:2020qqa,DeLuca:2020bjf} and reviewed for completeness in Appendix~\ref{app:B}. We highlight for clarity that a lower cut-off is associated to stronger accretion and vice-versa. Values around $z_\co \simeq 30$ corresponds to negligible accretion in the mass range of interest for the LVC observations.

As the spin of PBHs formed from the collapse of density perturbations in the early universe is expected to be negligible at formation~\cite{DeLuca:2019buf,Mirbabayi:2019uph}, a non-zero effective spin parameter 
$\chi_\text{\tiny eff}$
 can only be acquired by PBH binaries through an efficient phase of accretion~\cite{DeLuca:2020qqa,DeLuca:2020bjf}, see Appendix B.
 As such, a peculiar characteristics of the PBH model is the correlation between large values of binary total masses and large values of the spins. This correlation will be considered fully when computing the event parameter distributions in Eq.~\eqref{ppop}.
The PBH spin directions are not correlated and randomly distributed on the two-sphere.

We also highlight the enhancing contribution coming from the terms in the second line of Eq.~\eqref{PBHrate} which accounts for both the hardening of binaries and the reduction of merger timescale through GW emission due to mass accretion of the PBHs participating in the binaries.

Let us also stress the presence of the suppression factor $S\lp M^\ii_\text{\tiny tot}, f_\PBH,\psi  \rp \equiv S_1 \times S_2$, which accounts for both the suppression of binary formation due to the surrounding DM matter inhomogeneities (not in the form of PBHs) and the disruption of binaries due to neighbouring PBHs~\cite{Raidal:2018bbj,Vaskonen:2019jpv,Jedamzik:2020ypm,Young:2020scc,j2,DeLuca:2020jug,Tkachev:2020uin,Hutsi:2020sol}.
In particular, the second piece $S_2$ specifically accounts for  disruption of PBH binaries in early sub-structures and clusters throughout the the history of the universe as binaries typically form before matter-radiation equality. The two pieces read~\cite{Hutsi:2020sol}
\begin{align}
	S_1 (M_\text{\tiny tot}, f_\PBH, \psi)& \thickapprox 1.42 \llp \frac{\langle m^2 \rangle/\langle m\rangle^2}{\bar N(y) +C} + \frac{\sigma ^2_\text{\tiny M}}{f^2_\PBH}\rrp ^{-21/74} \exp \llp -  \bar N(y ) \rrp 
	\qquad 
	\text{with}
		\qquad 
		\bar N(y) \equiv \frac{M_\text{\tiny tot}}{\langle m \rangle } \frac{f_\PBH}{f_\PBH+ \sigma_\text{\tiny M}}
		\label{S1}
	\\
	S_2 (x) & \thickapprox \text{min} \llp 1, 9.6 \cdot 10^{-3} x ^{-0.65} \exp \lp 0.03 \ln^2 x \rp  \rrp 
	\qquad 	\qquad 	\,\,\ 
	\text{with}
		\qquad 
		x \equiv (t(z)/t_0)^{0.44} f_\PBH.
\end{align}
Notice also that, for $f_\PBH \lesssim 0.003$, one always finds $S_2\simeq 1$, i.e. the suppression of the merger rate due to disruption inside PBH clusters is negligible. This is also supported by the results obtained through a cosmological N-body simulation finding that PBHs are essentially isolated for a small enough abundance~\cite{inman}.
In Eq.~\eqref{S1} the constant $C$ is defined as in Eq.~(A.5) of Ref.~\cite{Hutsi:2020sol}.

\begin{figure}[t!]
	\centering
	\includegraphics[width=0.68 \linewidth]{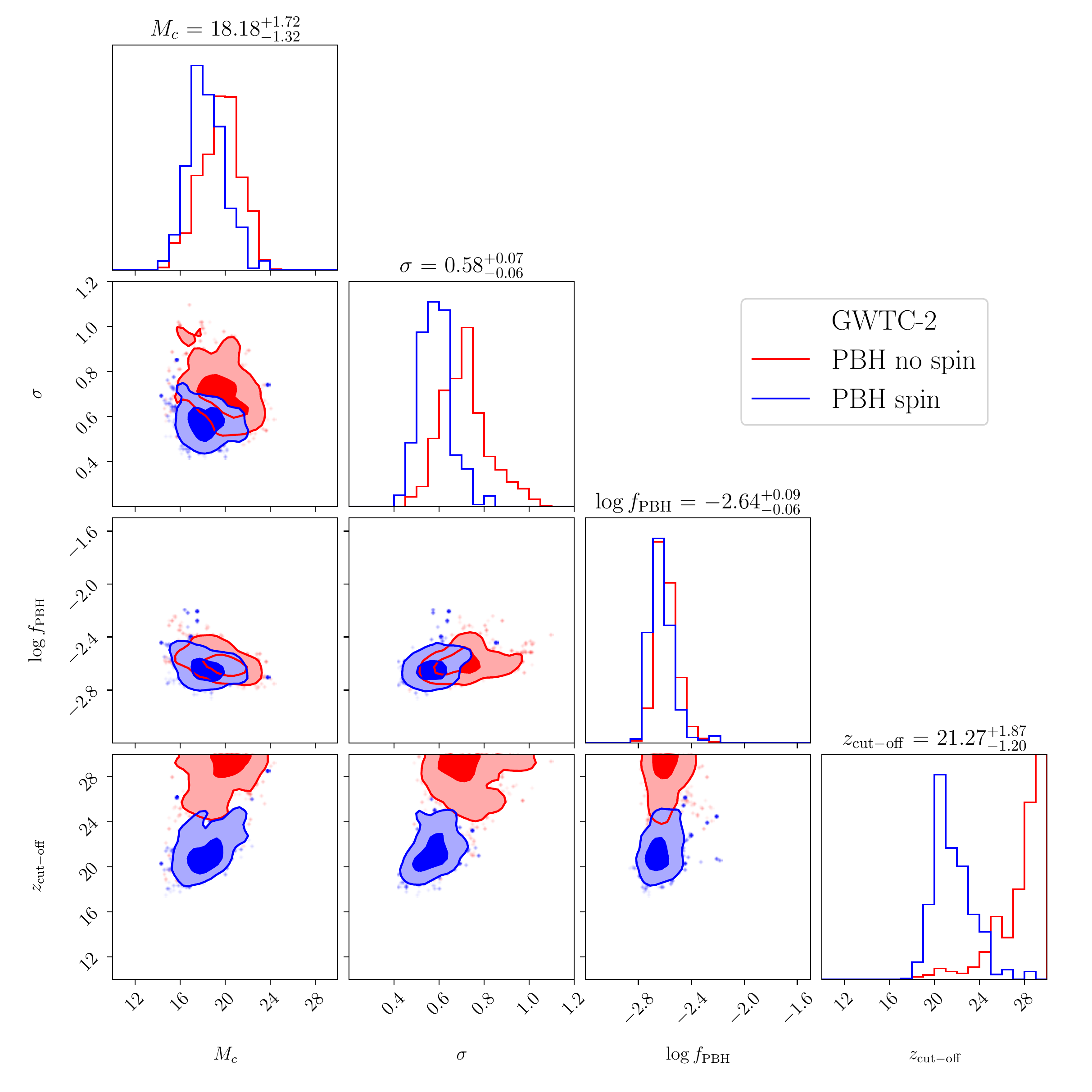}
	\caption{\it PBH population inference using the GWTC-2 catalog. Blue and red posteriors correspond to the case when 
the spin information is used in the inference or not respectively. This result reproduces the one published in  
Ref.~\cite{Wong:2020yig}.  The marginalised values on top of the plots reports the $90\%$ confidence interval.
	}
	\label{fig1}
\end{figure}

The hyperparameters of the PBH model are $\{ M_c, \sigma, f_\text{\tiny PBH}, z_\text{\tiny cut-off} \}$ and the priors 
chosen to perform the analysis are reported in Table~\ref{tab1}. 
In Fig.~\ref{fig1} we report the result of the Bayesian inference using GWTC-2 dataset and assuming all events have 
primordial origin. 
The result is consistent with what found in Ref.~\cite{Wong:2020yig} by means of machine learning empowered hierarchical 
Bayesian analysis and it is reported here in order to allow for a full comparison with the mixed ABH-PBH case.

\begin{table}
\renewcommand{\arraystretch}{1.3}
\caption{Prior ranges for the hyperparameters of the PBH model. We assume uniform distributions.}
\begin{tabularx}{\columnwidth}{Xcccc}
\hline
  \hline
Parameter&     
      \hspace{.3 cm}  $z_\text{\tiny cut-off}$  \hspace{.3 cm}   & 
       \hspace{.3 cm} $M_c [M_\odot]$  \hspace{.3 cm} &
        \hspace{.3 cm}  $\sigma$  \hspace{.3 cm} &
         \hspace{.3 cm} $\log f_\text{\tiny PBH}$  \hspace{.3 cm}  \\ 
Prior range & $[10,30]$  & 
           $[5,40]$ &
          $[0.1,1.1]$  & 
            $[-4,-1.5]$ \\
 \hline
  \hline
\end{tabularx}
\label{tab1}
\end{table}

The most striking feature of the posteriors for the full PBH analysis is given by the preference for a large cut-off redshift when no spin information is used. This is the result of the fact that, given accretion is a non-linear process which enhances the high mass tail of the binary mass function, lower cut-off are typically associated to a larger fraction of events with masses in the heavy portion of the observable window at odds with the dataset.
However, this tendency is inverted when the spin information is used, by the need to explain the ${\cal O}( 10) $ events in the GWTC-2 catalog with a spin incompatible with zero. Those can only be accommodated in the PBH model if some accretion is present. 

\section{The LVC astrophysical phenomenological model}
\renewcommand{\theequation}{4.\arabic{equation}}
\setcounter{equation}{0}
\label{sec3}

\noindent
We consider a representative astrophysical phenomenological model, physically equivalent to the  ``Truncated Model'' adopted by the LVC in the population analysis of both the GWTC-1~\cite{LIGOScientific:2018jsj} and GWTC-2~\cite{Abbott:2020gyp} catalogs.  
This model considers the local merger rate $R_0$, the mass spectrum slope $\zeta$, both the minimum $m_\text{\tiny min}$ and maximum  $m_\text{\tiny max}$ mass,  and the exponent $\beta$ of the symmetric mass ratio as free parameters. 
The corresponding  differential merger rate can be parametrised  as\footnote{Notice that we adopt a symmetric parametrisation of the merger rate with respect to the BH masses, differently from what the LVC collaboration used in Ref.~\cite{Abbott:2020gyp}.}
\begin{equation}\label{rateABH}
	\frac{\d R_\text{\tiny ABH}}{\d m_1\d m_2 } =  {\cal N} R_0  (1+z)^\kappa (m_1+m_2)^\alpha \left[ \frac{m_1 m_2 }{(m_1+m_2)^2}\right]^\beta \psi (m_1) \psi(m_2),
\end{equation}
where ${\cal N}$ is a normalization factor ensuring the local merger rate is $R(z=0) \equiv R_0$. Also, the mass function considered is a power law with sharp cut-off at both ends as
\begin{equation}
	\psi(m|\zeta,m_\text{\tiny min},m_\text{\tiny max}) \propto m^{-\zeta}
	\qquad \text{for} \qquad  
	m_\text{\tiny min} < m < m_\text{\tiny max}, 
\end{equation}
with an overall constant enforcing unitary normalization as 
$
\int \psi (m) \d \ln m =1$.

In the recent analysis by the LVC they adopted three different spin models named ``Default'', ``Gaussian''  and ``Multi-'' spin model, respectively, see Appendix~D of Ref.~\cite{Abbott:2020gyp} for more details. In this work we adopt the  ``Gaussian'' model and, as mentioned  in Sec.~\ref{sec1}, we neglect the precession spin parameter $\chi_\text{\tiny p}$ in the inference, since it is still poorly measured for most of the GW events~\cite{Abbott:2020niy}. In the ``Gaussian'' model the effective spin parameter $\chi_\text{\tiny eff}$ is distributed as a Gaussian function with mean $\mu_{\chi_\text{\tiny eff}}$ and variance $\sigma_{\chi_\text{\tiny eff}}$ truncated within the range $[-1,1]$. Notice that the population reconstructed using either the ``Gaussian'' and ``Default'' models are in agreement with each other in the analysis performed by the LVC~\cite{Abbott:2020gyp}.

The choice of the hyperparameters for the phenomenological ABH model is summarized in Table~\ref{tab2} along with the ranges for the priors considered in the analysis.
Notice that in the ABH model we fix the  evolution of the merger rate with redshift to the best fit value found in~\cite{Abbott:2020gyp}. We stress however that, being the data only limited to redshifts smaller than unity, the result of the inference is still largely insensitive to the merger rate evolution.

\begin{table}
\renewcommand{\arraystretch}{1.3}
\caption{Prior ranges for the hyperparameters of the ABH model. We assume uniform distributions.}
\begin{tabularx}{\columnwidth}{Xccccccccc}
\hline
  \hline
Parameter  &        \hspace{.25 cm} $R_0  [\text{Gpc}^{-3} \text{yr}^{-1}] $\hspace{.25 cm}   & 
  \hspace{.25 cm} $\kappa$ \hspace{.25 cm} & 
  \hspace{.25 cm} $\alpha$ \hspace{.25 cm} &
  \hspace{.25 cm} $\beta$ \hspace{.25 cm} & 
  \hspace{.25 cm} $\zeta$ \hspace{.25 cm} &
  \hspace{.25 cm} $m_\text{\tiny min} [M_\odot] $ \hspace{.25 cm} &
  \hspace{.25 cm}  $m_\text{\tiny max} [M_\odot] $ \hspace{.25 cm}   &
    \hspace{.25 cm}  $\mu_{\chi_\text{\tiny eff}}$ \hspace{.25 cm}   &
      \hspace{.25 cm}  $\sigma_{\chi_\text{\tiny eff}}$ \hspace{.25 cm}   
  \\
Prior range &  $[1,100]$ &  1.5      & 0  & [-4,12] & [-6,0]  & $[2,10]$ & $[30,100]$  & [-1,1] & [0,1]\\
 \hline
  \hline
\end{tabularx}
\label{tab2}
\end{table}
\begin{figure}[t!]
	\centering
	\includegraphics[width=0.88 \linewidth]{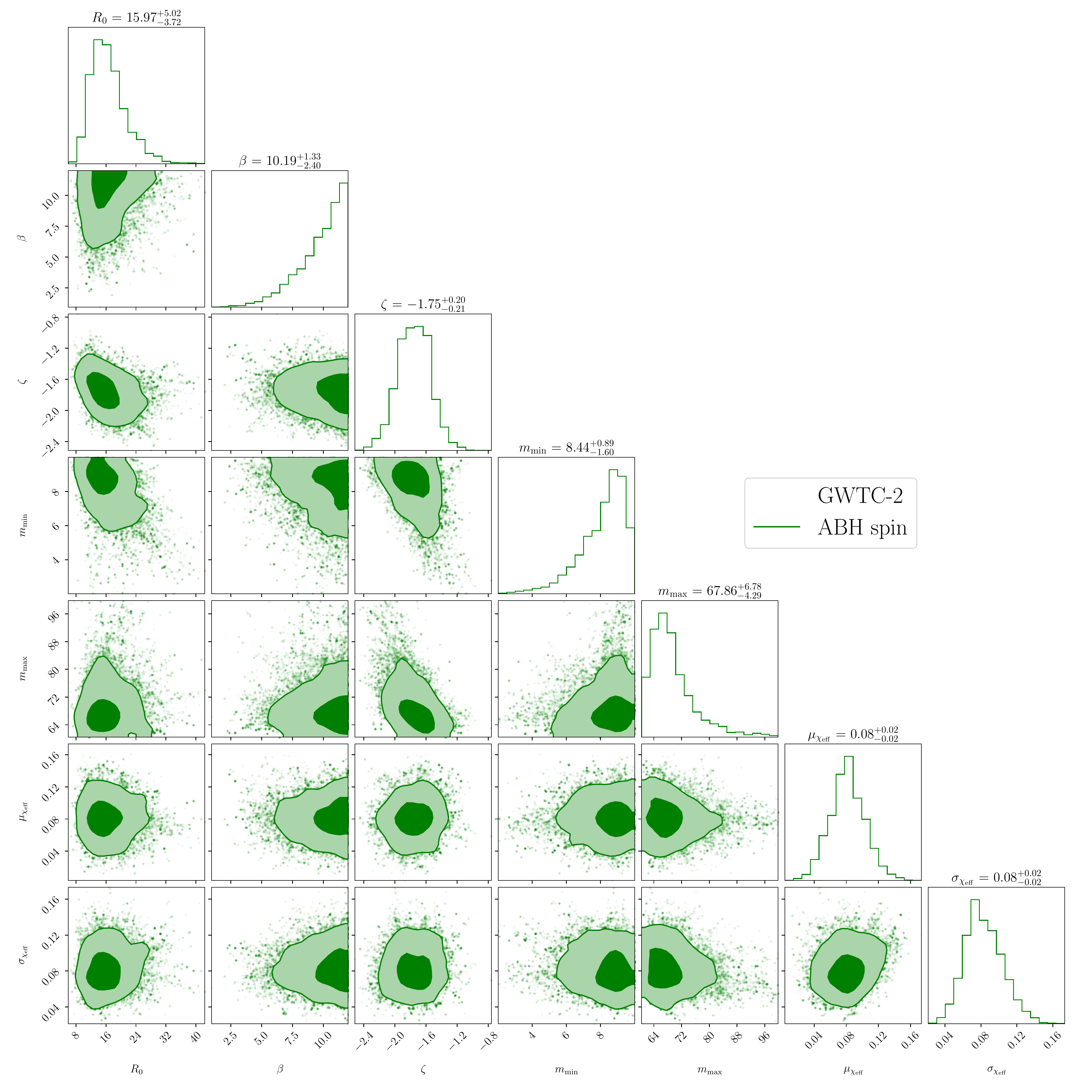}
	\caption{\it ABH population posterior found by running the inference on the GWTC-2 catalog.  
	The marginalised values on top of the plots report the $90\%$ confidence interval.
	}
	\label{fig2}
\end{figure}
In this section, we consider the scenario in which all events in the GWTC-2 catalog (as specified in Sec.~II) are interpreted as ABH binaries and the resulting posterior distribution is shown in Fig.~\ref{fig2}.
Being the overall rate consistent with the one inferred by the LVC, $R_0 = 15.97^{+5.02}_{-3.72} / \text{Gpc}^3\text{yr}$ , one relevant feature of the posterior distribution is its sharp drop of the posterior for values of $m_\text{\tiny max}$ smaller than $\lesssim 60 M_\odot$. This is due to the lack of support of the model where the posterior of the primary mass of the mass gap event GW190521 falls~\cite{massgapevent}. This also follows from the choice of a hard cut-off at the maximum mass in the ABH model.  
An analogous behaviour is observed for $m_\text{\tiny min}$.
It is interesting to notice the negative correlation between $R_0$ and $\zeta$. This can be explained as follows. With a steeper mass function the binary masses are expected to be on the lighter portion of the observable mass range. However, having a smaller SNR, lighter binaries are harder to observe. Therefore, the rate must be enhanced in order to predict enough observable events for the model to be compatible with the GWTC-2 catalog. A similar correlation between the slope $\zeta$ and the maximum mass is observed since if the mass function extends to larger masses, the posterior moves to steeper profiles in order to reduce the contribution of massive events in the observable range.

Furthermore, the preference for a large value of $\beta$  signals a tendency towards symmetric binaries. This is intuitively understood by noticing that the vast majority of events in the GWTC-2 catalog is compatible with $q \sim 1$. This feature was already observed using the GWTC-1 catalog. 
Finally, the effective spin parameter distribution is found to be narrow and a slight offset towards positive $\chi_\text{\tiny eff}$. 
Even though there is no correlation between the spin distribution and the mass distribution in the ABH merger rate model in Eq.~\eqref{rateABH}, 
there exists a well-known parameter degeneracy between $(\chi_\text{\tiny eff},q)$ in the GW data~\cite{deg1,deg2,deg3,deg4,deg5,deg6}. We checked that the reconstructed population parameters are only slightly affected by the introduction of spin information in the inference in this case.

\section{Mixed PBH + ABH scenario}
\renewcommand{\theequation}{5.\arabic{equation}}
\setcounter{equation}{0}
\label{sec4}

\noindent
In this section we investigate the mixed scenario where the GWTC-2 population is supposed to contain both ABH and PBH binaries and
perform the inference accounting for possible contributions to the observed binaries from both models.
Similar analysis accounting for multiple astrophysical formation channels, but
without accounting for a possible PBH contribution, can be found in Refs.~\cite{Zevin:2020gbd,Wong:2020ise,Ng:2020qpk}.

\begin{figure}[t!]
	\centering
	\includegraphics[width=1 \linewidth]{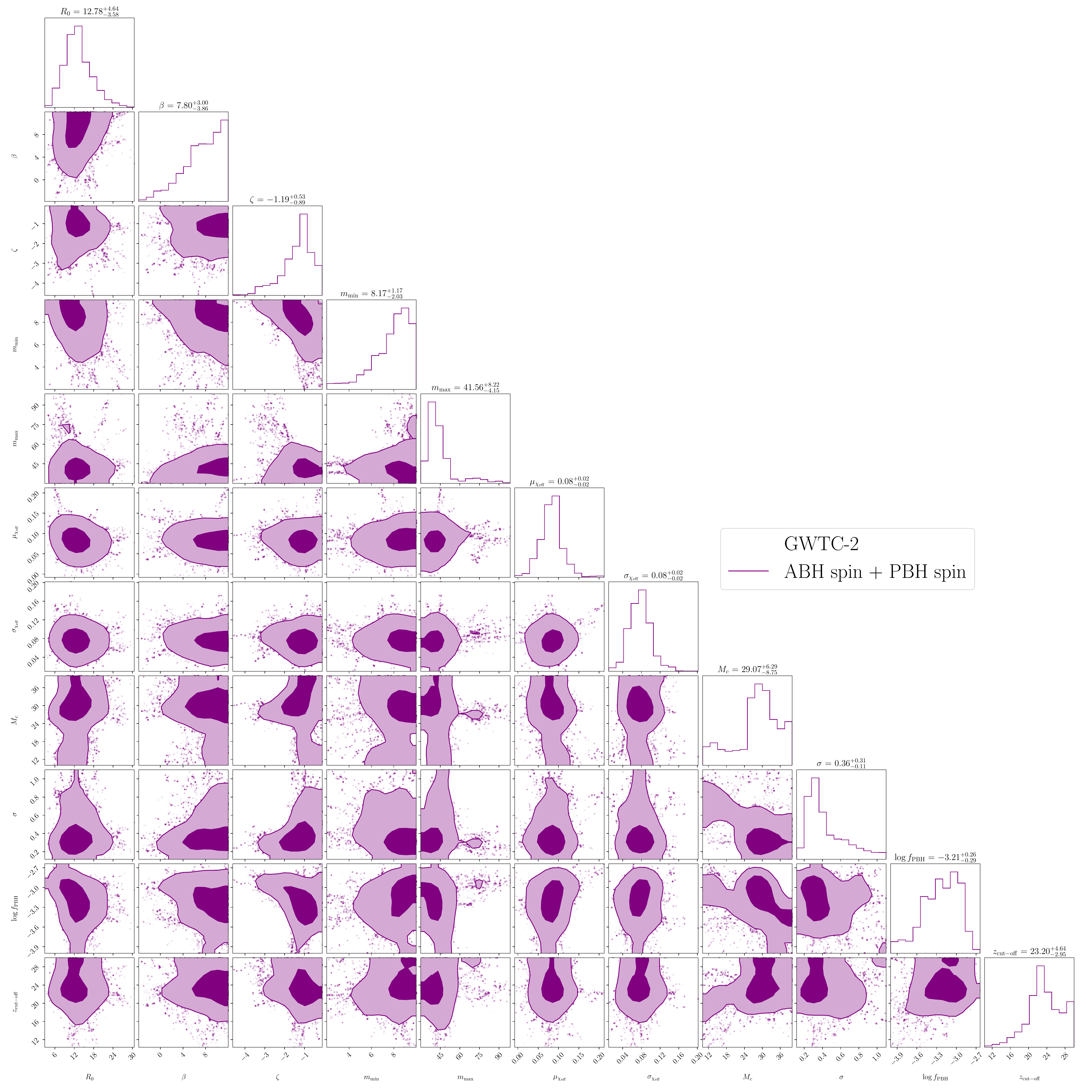}
	\caption{\it Full PBH+ABH population inference using the GWTC-2 catalog. 
	The marginalised values on top of the plots reports the $90\%$ confidence interval.
	}
	\label{fig6}
\end{figure}

Consistently with the previous sections, the set of hyperparameters of the mixed model contains $7$ astrophysical inputs $\{R_0, \beta, \zeta, m_\text{\tiny min},  m_\text{\tiny max} , \mu_{\chi_\text{\tiny eff}}, \sigma_{\chi_\text{\tiny eff}}\}$ as well as $4$ primordial parameters $\{ M_c, \sigma, f_\PBH, z_\co\}$. 
The result of the inference is shown in Fig.~\ref{fig6}.

\begin{figure}[t!]
	\centering
	\includegraphics[width=0.49 \linewidth]{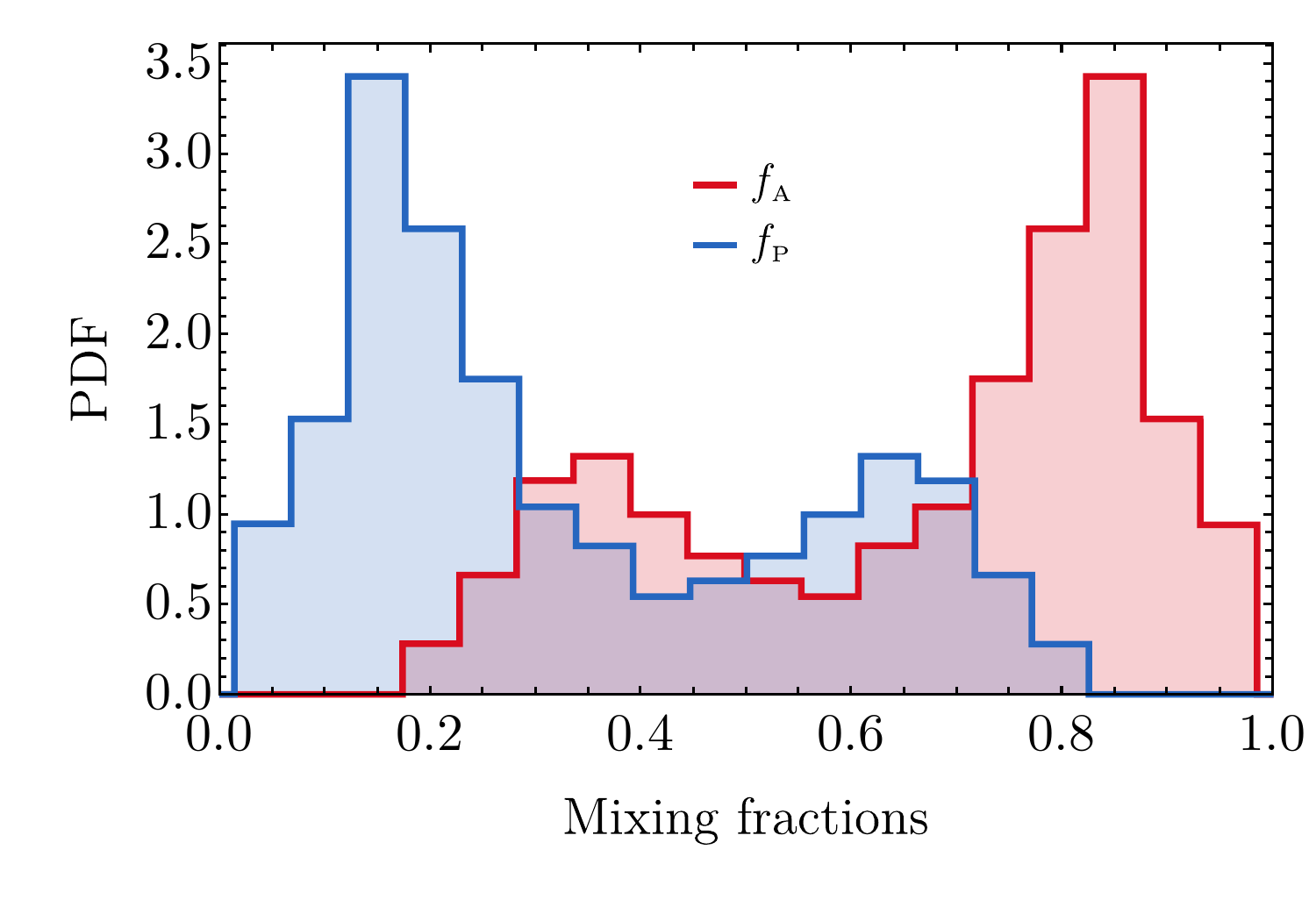}	
	\caption{\it Distribution of the mixing fraction between ABHs and PBHs as found by a Bayesian inference on the GWTC-2 catalog with the mixed ABH-PBH model.}
	\label{fig8}
\end{figure}

The first noticeable feature  is the reduction of the values of both  $f_\PBH$ and $R_0$ in the mixed scenario compared to the corresponding single-population scenarios studied in the previous sections. As both parameters are directly related to the overall merger rate in both the ABH and PBH sectors respectively, their values are reduced as only a smaller portion of events can be ascribed to either sectors. In particular one can compute the mixing fractions in the GWTC-2 events, defined as 
\begin{align}
f_\text{\tiny P} &\equiv N^\text{\tiny det}_\text{\tiny PBH}/(N^\text{\tiny det}_\text{\tiny ABH}+ N^\text{\tiny det}_\PBH).
\\
f_\text{\tiny A} &\equiv N^\text{\tiny det}_\text{\tiny ABH}/(N^\text{\tiny det}_\text{\tiny ABH}+ N^\text{\tiny det}_\PBH) \equiv 1-f_\text{\tiny P}.
\end{align}
In Fig.~\ref{fig8} we show both the distributions of  
the mixing fractions. More events are interpreted as coming from the ABH sector with respect to the PBH one, having a fraction of events peaking at around $f_\text{\tiny A} \simeq 0.8$, corresponding to about four times the number of PBHs in the GWTC-2 catalog. This indicates a preference of the data for having a larger contribution coming from the ABH phenomenological model with only a smaller contribution to the GWTC-2 catalog from PBH binaries.
It is important to stress though that the posterior distribution for the mixing fraction of PBH $f_\text{\tiny P}$ is incompatible with zero.
The secondary peak structure in $f_\text{\tiny P} \simeq 2/3 $ is due to a subdominant region in the posterior with small $m_\text{\tiny max}$ along with large $f_\text{\tiny PBH}$. Overall, there is a statistically significant portion of the posterior for which the mixing fractions are comparable.

We also perform a comparison of how well the single-population (either ABH or PBH) scenario is able to explain the 
entirety of GW events detected so far as compared to our best-fit ABH+PBH mixed model.
In Refs.~\cite{Hall:2020daa} and~\cite{Hutsi:2020sol} using the GWTC-1 and GWTC-2 dataset respectively, it was reported
that the LVC phenomenological model was strongly favoured when compared to the PBH scenario. 
This conclusion is confirmed by our analysis, as 
the Bayes factor (see Table~\ref{tabbayes}) turns out to be $\log_{10}{\cal B}^\ABH_\PBH \approx 2.74$, which implies a decisive preference for the ABH model. This shows that  the inclusion of accretion in the PBH model, as well as spin information in the inference, is not able to reduce the gap between the ABH and PBH models in the attempt to explain the LVC observation by means of a single population.
However, we find that $\log_{10}{\cal B}^{\ABH+\PBH}_\ABH \approx 2.17$, providing decisive 
evidence (according to Jeffreys' criterion~\cite{Jeffreys}) for our best-fit mixed ABH-PBH model relative to a 
single-population ABH model.
This implies  that the presence of some PBH events in the GWTC-2 catalog seems to be demanded to better fit the data. 
\begin{table}
\renewcommand{\arraystretch}{1.3}
\caption{Bayesian evidence ratios.}
\begin{tabularx}{\columnwidth}{Xcccc}
\hline
  \hline
Model ${\cal M}$&     
      \hspace{.5 cm}  PBH  \hspace{.5 cm}   & 
       \hspace{.5 cm} ABH  \hspace{.5 cm} &
        \hspace{.5 cm}  PBH+ABH  \hspace{.5 cm} 
        \\ 

$\log_{10} {\cal B}^{\cal  M}_{\PBH + \ABH} \equiv \log_{10} \lp Z_{\cal  M}/Z_{\PBH + \ABH} \rp $ &  -4.91 & 
           -2.17 &
          0 \\
 \hline
  \hline
\end{tabularx}
\label{tabbayes}
\end{table}

\begin{figure}[t!]
	\centering
	\includegraphics[width=0.99 \linewidth]{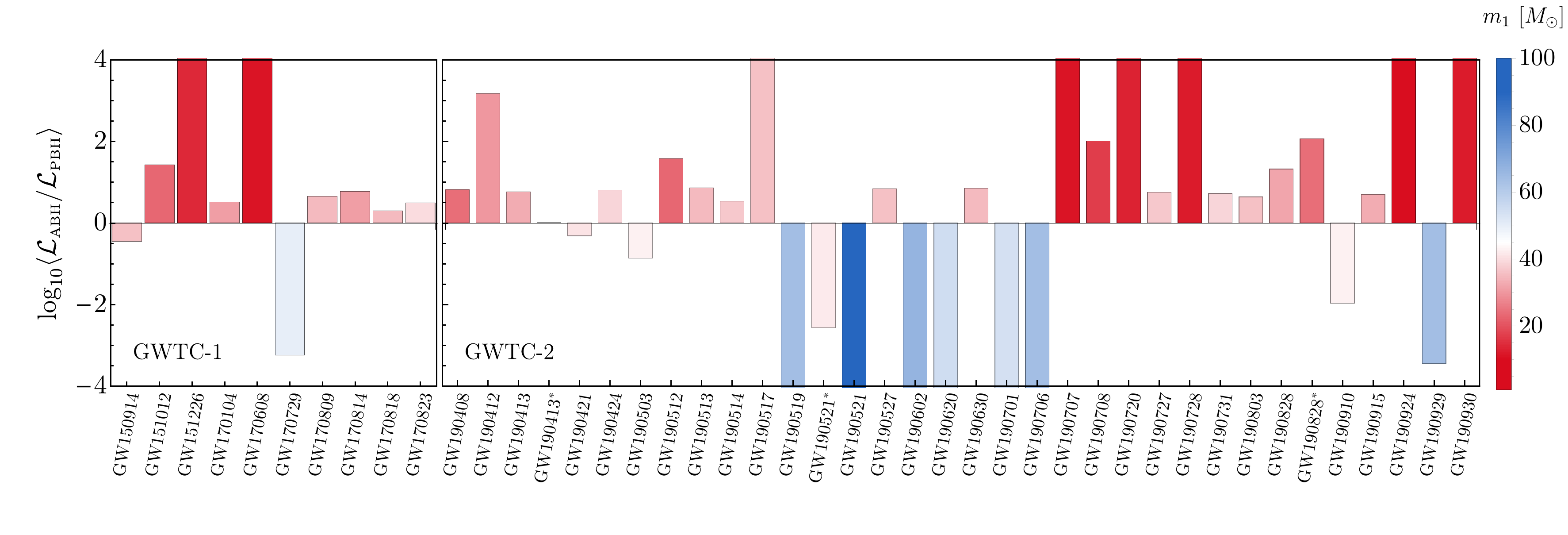}
	\caption{\it 
Plot of the ratio between  the ABH and PBH likelihood for each event in the GWTC-2 catalog. The ratio is averaged over 
the population posterior in Fig.~\ref{fig6}. This plot indicates that the inference confidently identifies massive 
events in the GWTC-2 catalog as PBHs.
The labels with an asterisk correspond to GW190413*=GW190413\_134308,  GW190521*=GW190521\_074359 and GW190828*=GW190828\_065509
respectively. 
}
	\label{fig7a}
\end{figure}

\begin{figure}[t!]
	\centering
	\includegraphics[width=0.49 \linewidth]{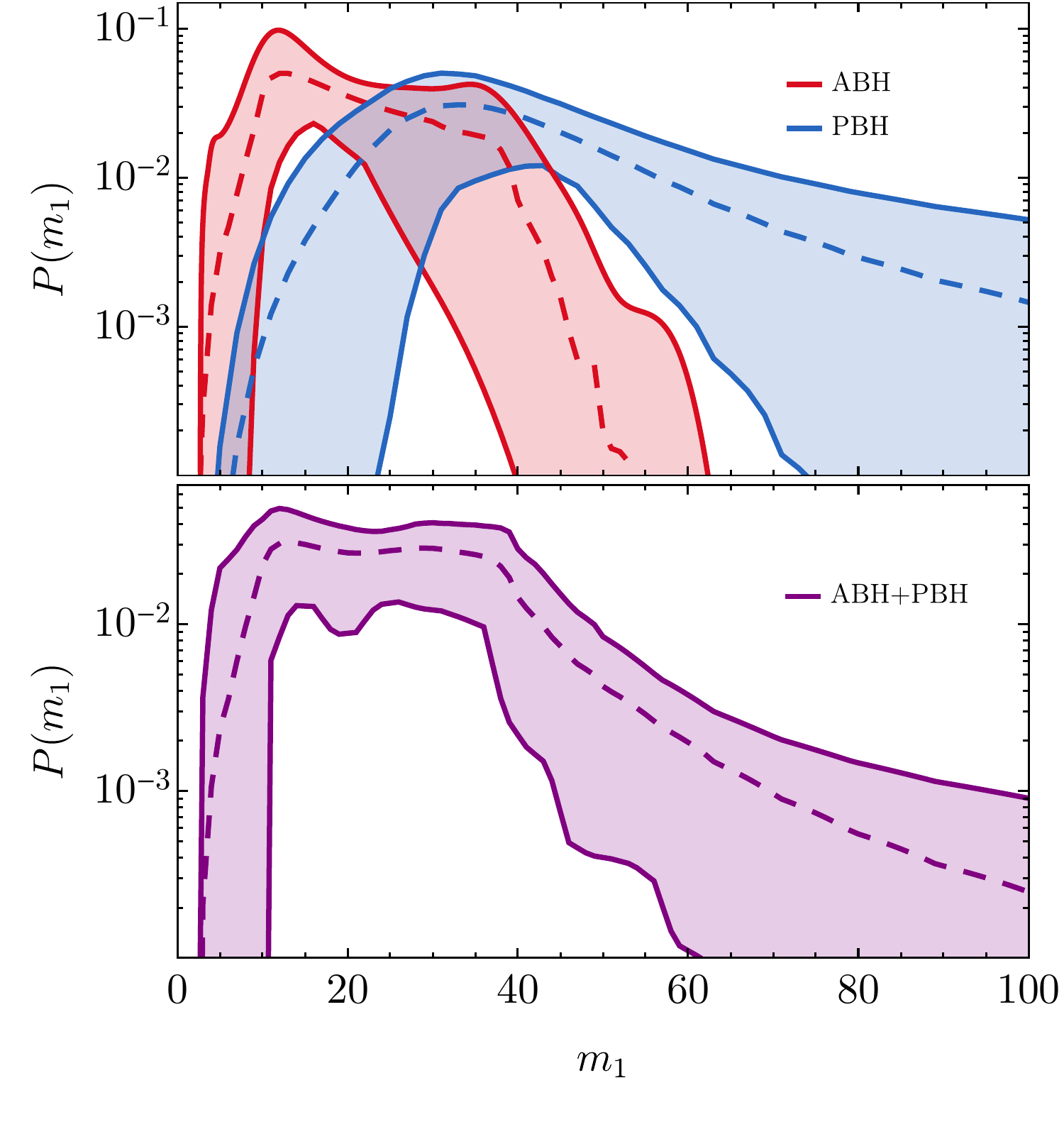}
	\includegraphics[width=0.49 \linewidth]{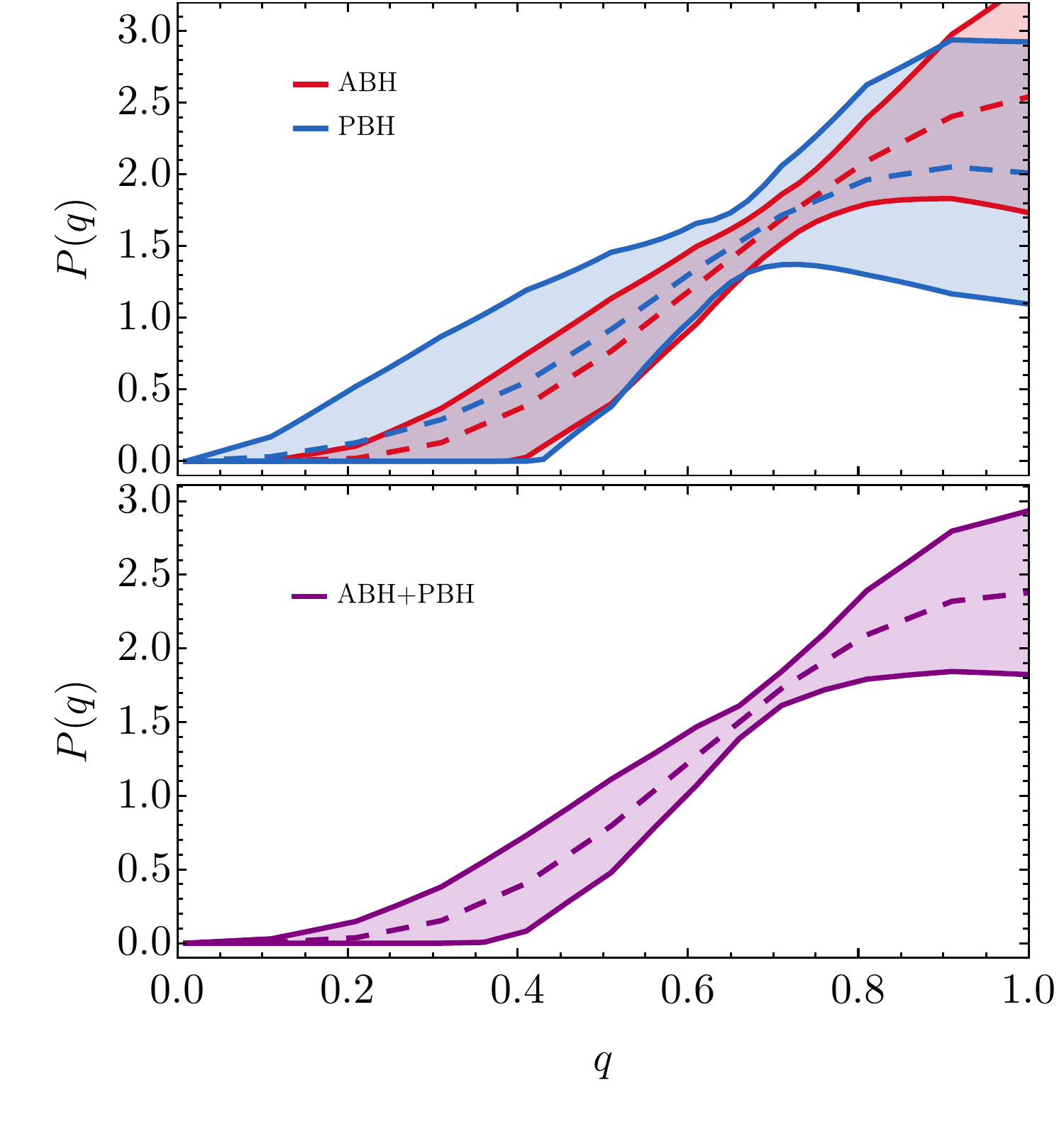}
	\caption{\it \textbf{  Top:} Contribution to the primary mass and mass ratio distributions for each sectors separately.
	\textbf{ Bottom:}
	 Cumulative population distribution.
	Notice that the relative contribution of each population to the overall event parameter distribution is weighted by the overall relative number of events expected from each population as shown in Fig.~\ref{fig8}. The bands around the dashed lines represent the $90\%$ confidence spread.}
	\label{fig7}
\end{figure}

Overall, the mass distributions found for both models are generally similar to the one observed in the isolated 
scenarios. 
One important exception is provided by the preferred value of the upper mass cut-off $m_\text{\tiny max}$ in the ABH sector. Indeed, compared to the single-population scenario, the upper mass bound is found to be significantly smaller being $m_\text{\tiny max} = 41.6^{+8.2}_{-4.2} M_\odot$. This is due to the strong preference of the inference to interpret  massive events as primordial binaries, see Figs.~\ref{fig7a} and \ref{fig7} and their discussion below. As a consequence, also the mass function is found to be slightly less steep, due to the negative correlation between $(m_\text{\tiny max},\zeta)$ already discussed in the previous section.
In the PBH sector, one relevant  difference is in the reduced value of $f_\PBH$, which confirms that only a subdominant portion of the DM is allowed to be in the form of PBHs.
Also, the preferred PBH mass function is shifted towards larger masses (with a larger reference mass scale $M_c$) and a narrower shape, as it only needs to accomodate a smaller portion of the GWTC-2 events compared to the single-population scenario.
Finally, the mixed population inference confirms a preference for weak accretion with a value of $z_\co$ which is compatible with the one found in Sec.~\ref{sec2}. The posterior on $z_\co$ is broader than in the single-population PBH case, since in the mixed scenarios PBHs are subdominant and the hyperparameters of their population are less constrained by the data.

In order to guide the understanding of how the GWTC-2 events are interpreted by the inference, in 
Fig.~\ref{fig7a} we show the ratio of the contribution to the likelihood function ${\cal L} \equiv p({{\bm \lambda}|{\bm 
d}})/\pi({\bm \lambda})$ in Eq.~\eqref{eq:HBA}  from both the astrophysical and primordial sectors, averaged over the 
population posterior. The plot clearly shows that events with primary masses which have support  in the astrophysical 
mass gap (i.e. above ${\cal O}(45) M_\odot$) are confidently assigned to the PBH population.

The same information is contained in Fig.~\ref{fig7}, where we show the distribution of primary mass $m_1$ and mass ratio $q$ for both cases in both sectors separately and combined. 
The most striking difference between the two sectors is the presence of a tail at large masses well within the upper mass gap in the PBH binary $m_1$ distribution. On the other hand, the distribution of the mass ratio is found to be similar in both sectors, with a slightly enhanced ability of the PBH model to accomodate small mass ratios. 
We again conclude that the fraction of PBH binaries in a mixed model could be responsible for events in the GWTC-2 catalog with large masses which do not fit well within the ABH model.

It is interesting to compare our findings to the population analysis performed by the LVC collaboration. As also found in this work, Ref.~\cite{Abbott:2020gyp} concluded  that events in the GWTC-2    catalog are suggesting the presence of an additional distinct population of binaries with primary masses above $\thickapprox 45 M_\odot$. Within the standard stellar formation scenario, it is  difficult to accomodate massive events due to the Pulsational Pair Supernova Instability (PPSN) preventing the formation of binaries with masses above $\thickapprox 45 M_\odot$, even though the precise location of such a cut-off is still uncertain \cite{rakavy,Barkat:1967zz,fraley,Heger:2001cd,Woosley:2007qp,Belczynski:2016jno,Woosley:2016hmi,Stevenson:2019rcw,Farmer:2019jed,Renzo:2020rzx,Mapelli:2019ipt,Croon:2020oga,Marchant:2020haw,Ziegler:2020klg,Belczynski:2020bca}.  One interesting possibility within the astrophysical sector is that massive events in the catalog are coming from second generation mergers in globular clusters or galactic nuclei \cite{Fishbach:2017dwv,Gerosa:2019zmo,Rodriguez:2019huv,Baibhav:2020xdf,Kimball:2020opk,Samsing:2020qqd,Mapelli:2020xeq}.
Our findings show that introducing a PBH population of binaries in the inference naturally leads to the interpretation of those events as coming from  primordial binaries. Interestingly, the distribution of mass ratio reconstructed from the data is found to be similar to the astrophysical one (i.e. both showing the same tendency for symmetric binaries), in contrast to the expectation of second generation mergers which predicts a smaller mass ratio (see for example Ref.~\cite{Gerosa:2017kvu}). 
This is quite an interesting  difference, which could be used in the future to disentangle primordial and second generation mergers in the high-mass portion of the dataset.

\section{Implications for 3G detectors} 
\renewcommand{\theequation}{6.\arabic{equation}}
\setcounter{equation}{0}
\label{sec5}

\noindent 
One of the most prominent differences between the PBH scenario and any astrophysical formation scenario is the 
predicted redshift evolution of the merger rate. Even though the GWTC-2 catalog is still limited to ``local'' GW signals 
coming from sources at redshift up to $z \lesssim1$, future observations with larger horizons will allow to use the 
merger rate evolution to distinguish between different binary BH populations at 
current~\cite{Fishbach:2018edt,Callister:2020arv,Fishbach:2021yvy} and future 
detectors~\cite{Ng:2020qpk,Vitale:2018yhm,Ding:2020nll,Chen:2019irf}.

The redshift evolution of the merger rate density for the PBH model is found to be monotonically increasing as~\cite{Raidal:2018bbj,DeLuca:2020qqa}
\begin{equation}\label{redevo}
 R_\PBH (z)  \thickapprox t^{-34/37}(z), 
\end{equation}
extending up to redshifts $z={\cal O}(10^3)$. Notice that the evolution of the merger rate with time shown in Eq.~\eqref{redevo} is entirely dictated by the dynamics of primordial binary formation (i.e. how pairs of PBHs decouple from the Hubble flow) before the matter-radiation equality era. As such, it is a robust prediction within the PBH scenario.
On the other hand, the merger rate of ABHs is expected to peak at 
redshift of a few
with a possible second peak coming from a Pop III population at redshift $z\sim 10$~\cite{Ng:2020qpk,Kinugawa:2014zha,Kinugawa:2015nla,Hartwig:2016nde}.
In particular, astrophysical models predicts that BHs from Pop~III stars~\cite{Schneider:1999us,Schneider:2001bu,Schneider2003,Liu:2020ufc} should 
form at $z\sim 25$~\cite{Valiante2021} and one could conservatively assume that they form up to $z=30$. Their merger 
time depends on the formation mechanisms of the binaries and could range from ${\cal O}({\rm Gyr})$ (in which case they 
merge at $z\lesssim6$) to ${\cal O}(10\,{\rm Myr})$ if they form dynamically in Pop~III clusters, in which case they would merge 
almost at the redshift of BH formation~\cite{Valiante2021}.

Therefore, assuming the most conservative scenario (BH mergers formed from Pop~III clusters), the detection of a binary 
BH at redshift higher than $z\sim 30$ would be a smoking-gun signal in favour of PBHs, as no ABHs are expected at 
those or higher redshifts in a standard cosmology~\cite{Koushiappas:2017kqm}.
It is interesting therefore to make a forecast for the observability of possible PBH at high redshifts through 3G detectors by assuming the PBH mass function providing the subpopulation of the currently available GWTC-2 dataset found in this paper.

A description of the sensitivity of 3G detectors is reported in Appendix~\ref{app:A}.
In Fig.~\ref{fig11} we show the predicted distribution of observable events per year for ET and CE as a 
function of redshift, assuming the PBH population identified by the mixed population inference on the GWTC-2 catalog 
found in the previous section.
The ET telescope, at design configuration, will be able to observe at least one PBH event per year up to $z\sim 50$. A similar figure is found for CE.
In the right panel of Fig.~\ref{fig11} we also show the distribution as a function of the primary mass. As expected by 
looking at the observable horizon shown in Fig.~\ref{psd}, most of the distant binaries will have their primary mass 
around the characteristic PBH mass function scale $M_c$.
\begin{figure}[t!]
	\centering
	\includegraphics[width=0.488 \linewidth]{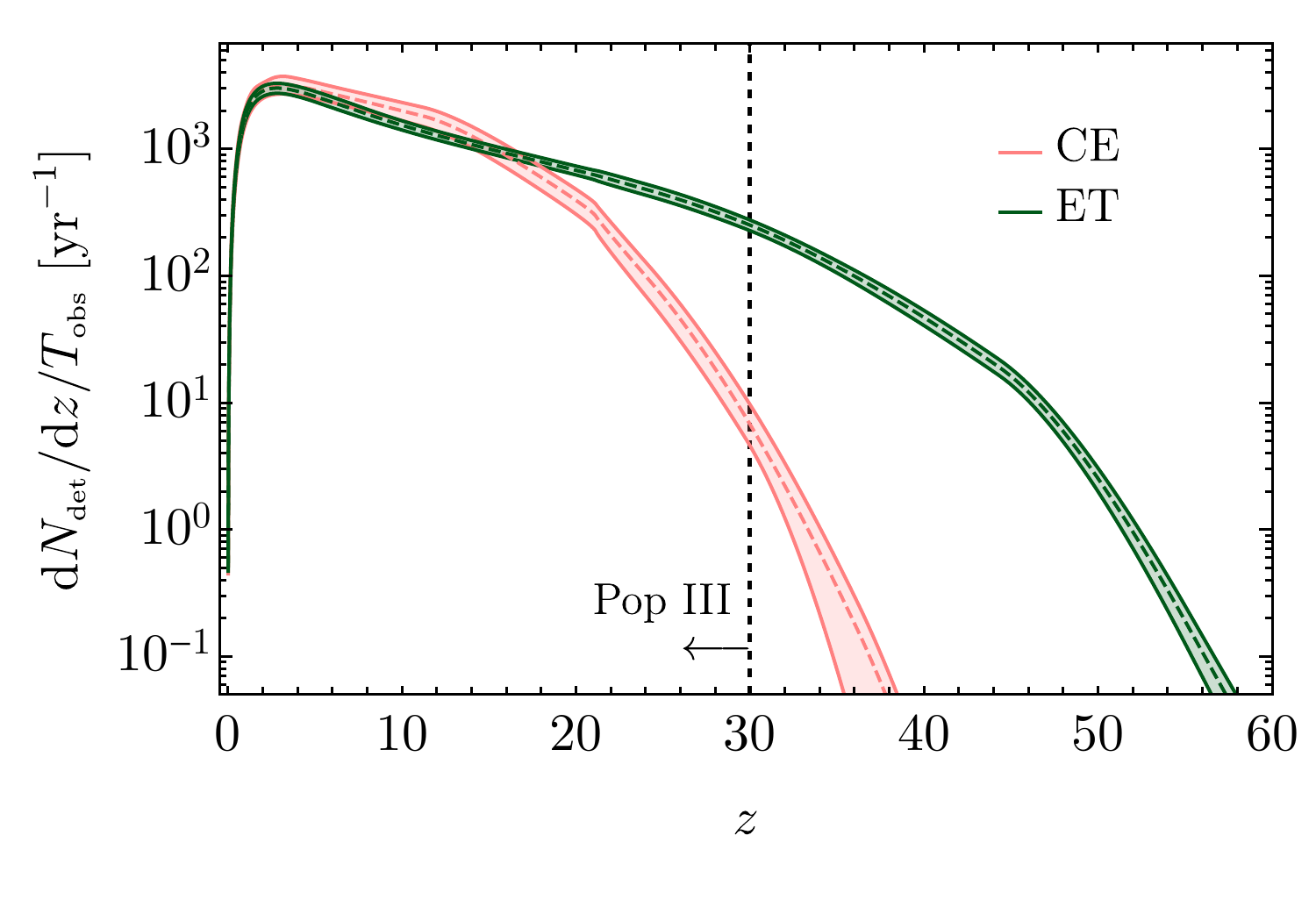}
	\includegraphics[width=0.502 \linewidth]{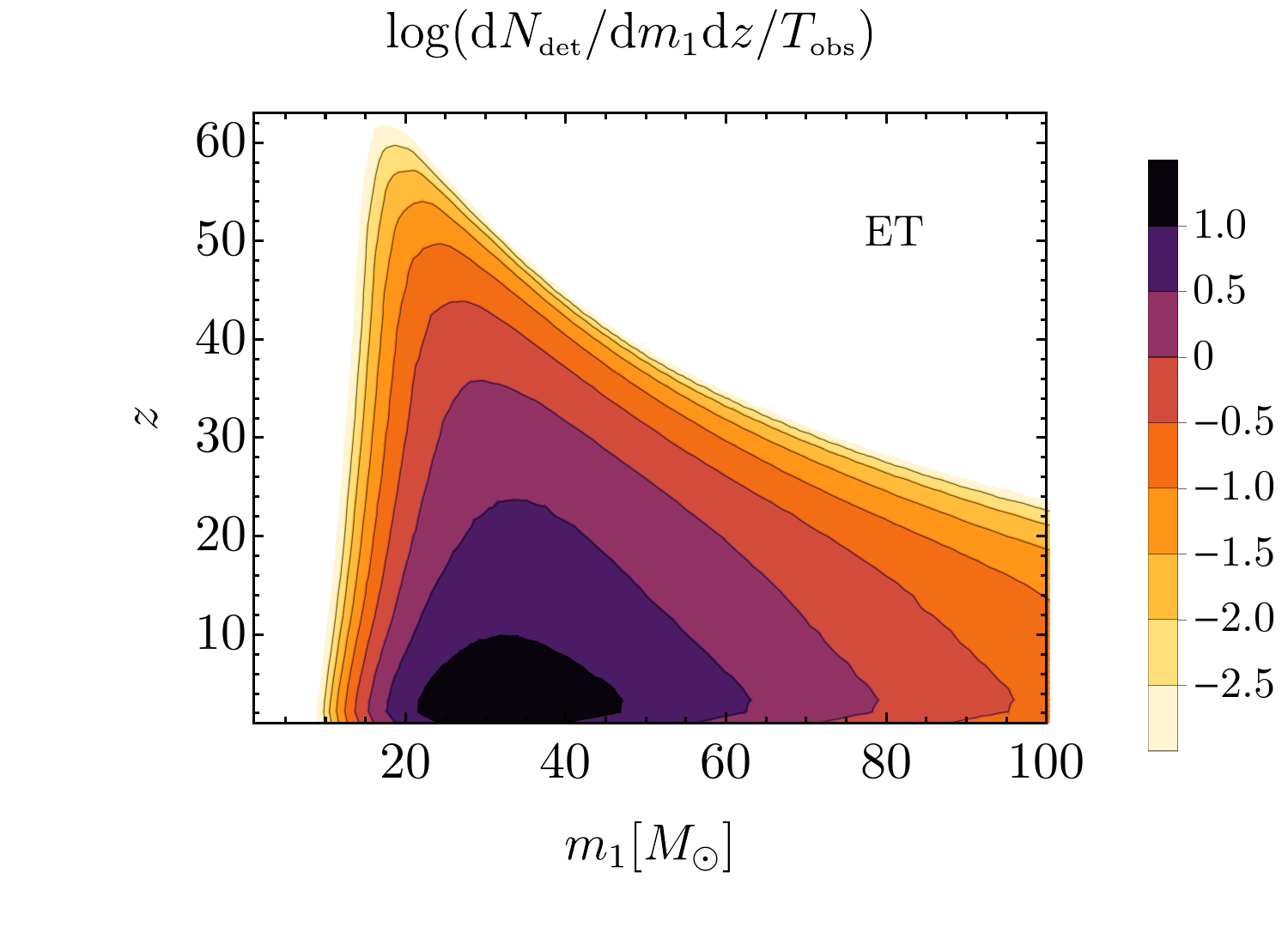}
	\caption{\it \textbf{  Left:} Distribution of observable events per year as a function of redshift at ET and CE coming from the PBH subpopulation. 
	\textbf{  Right:}  Distribution of observable events as a function of both primary mass and redshift at ET coming from the subpopulation of PBHs.}
	\label{fig11}
\end{figure}

Finally, we also computed the total number of observable distant events (i.e. with $z>30$) per year for both ET and CE detectors, finding 
\begin{align}
N^{\text{\tiny ET}}_\text{\tiny det}(z>30) = 1315 ^{+305}_{-168} \,/ \text{yr},
\qquad \qquad 
N^{\text{\tiny CE}}_\text{\tiny det}(z>30) = 12^{+22}_{-11} \,/ \text{yr}.
\end{align}
Note that the larger event rate expected for ET is due to the superior behaviour of the adopted ET-D sensitivity curve at low-frequencies compared to the CE phase-1 design (see Fig.~\ref{psd}), which is particularly relevant for heavy and/or high-redshift mergers. We conclude that 3G detectors have the potential of unequivocally confirm in the future the presence of a PBH subpopulation in the currently available dataset and to unveil a novel (primordial) family of BBHs. However, this conclusion depends also on the measurement accuracy of the redshift, which can be low for the most distant events even in the 3G era~\cite{Ng:2020qpk}.

One might think that a more conservative way of proceeding to estimate the maximum number of events for the future 3G detectors 
would be to maximise the PBH merger rate adopting the largest value of $f_\PBH$ currently allowed by the constraints. This would, in principle, allow a forecast independent from the details of the astrophysical model used to analyse the GWTC-2 data. However, within the PBH merger rate there exists a degeneracy in the choice of the mass distribution hyperparameters: the constraints on $f_\PBH$ drastically change when moving in the $(M_c,\sigma)$ plane. For instance, increasing $\sigma$ leads to much smaller values of $f_\PBH$ allowed by the CMB constraints (see for instance Refs.~\cite{DeLuca:2020qqa,Wong:2020yig}).

\section{Conclusions} 
\renewcommand{\theequation}{7.\arabic{equation}}
\setcounter{equation}{0}
\label{sec:conclusions}

\noindent 
In this work we have presented a hierarchical Bayesian investigation of the merger events in the GWTC-2 catalog with the goal of
understanding the nature of the BHs, in particular if they are of mixed astrophysical and primordial origin.
Using the LVC ``truncated'' phenomenological model for a single ABH population~\cite{LIGOScientific:2018jsj,Abbott:2020gyp}, the mixing fraction parameter inferred from the Bayesian analysis indicates that roughly $80\%$  of the BH mergers in the GWTC-2 catalog are of astrophysical origin. Furthermore, our best-fit mixed ABH+PBH model has a decisive statistical evidence relative to the single-population phenomenological ABH model, strongly supporting the existence of extra features which are not captured by the simple truncated ABH model. This suggests the existence of at least two distinct populations of BH mergers in LVC data and is compatible with a subdominant population of PBHs. The fraction of putative PBH events could in particular explain the events that are at odds with the standard astrophysical scenarios, such as those with binary components in the mass gap. Based on our best-fit mixed model, we predicted that the next-generations ground-based GW interferometers ET and CE could detect up to hundreds of BH merger at redshift $z\gtrsim30$, which can therefore be confidently identified as being primordial, provided their merger redshift can be measured with sufficient accuracy. The number of possible detections at $z\gtrsim30$ depends significantly on the low-frequency end of the 3G sensitivity curve. Thus, our analysis suggests that the low-frequency response of 3G detectors will be crucial for an important and unique science case~\cite{Maggiore:2019uih} such as the unambiguous detection of PBHs.

Our findings can be improved along several lines. First, on the PBH side, one might consider a mass function different from the adopted lognormal and improve on the accretion model. Secondly, on the astrophysical side, one might use more sophisticated and motivated ABH models. The simplest extension of our work in this respect would be to include a PBH population to the LVC ``power law+peak'' model used for the GWTC-2 catalogue~\cite{Abbott:2020gyp}. However, being the latter a phenomenological model that attempts to fit the data without any astrophysical input, it would be hard to disentangle its phenomenological features (in particular its Gaussian peak) from putative physical effects such as those coming from a subdominant PBH population. Notice, however,  that the mixed population  model described in this paper has a Bayesian evidence with respect to the single ABH model comparable to the  Bayesian evidence of the ``power law+peak" model with respect to the ``truncated" model adopted in Ref. ~\cite{Abbott:2020gyp}. In other words, PBHs seem to add the same features needed to match the phenomenological astrophysical model preferred by the LIGO/Virgo analysis. 

Another, and possibly more  interesting, avenue is to consider astrophysically-motivated ABH populations, for instance by accounting for both isolated and dynamical\footnote{In this context it is also interesting to note that a very recent analysis argues that the LVC phenomenological model could be entirely explained by dynamically-formed mergers in globular clusters~\cite{Rodriguez:2021nwd}.} formation channels~\cite{Wong:2020ise,Ng:2020qpk}, as well as including the merger rates from Pop~III binaries, which are relevant for forecasts with 3G detectors~\cite{Ng:2020qpk,Valiante2021}. Work along this line is in progress and will be reported elsewhere~\cite{inprep2}.

 In this paper we focused on confirming  the origin of BH mergers 
using the difference between the number of events expected at high
redshift between the ABH and PBH scenarios.
As a final remark, let us mention that a different possibility could be to use instead the clustering properties of mergers.
As the linear bias of ABH mergers is predicted to differ from the one of PBH mergers, it was proposed to distinguish between the
two populations by cross-correlating the GW event catalogue or a stochastic GW background with the galaxy catalogue~\cite{Namikawa:2016edr,Raccanelli:2016cud,Scelfo:2018sny,Canas-Herrera:2019npr,Calore:2020bpd}.
These different approaches are currently under intense study by various groups, and in the future it would be interesting to compare them.

\vskip 0.3cm
\noindent
\section{Acknowledgments}
\noindent
We thank A.~Maselli and  R.~Schneider for insightful discussions and we are indebted to V.~Baibhav,  E.~Berti, A.S. Biscoveanu, M. Maggiore, K. K. Y. Ng, F.~Pannarale, S. Vitale, and K.~W.~K.~Wong for discussions and comments on the draft.
Computations were performed at University of Geneva on the Baobab/Yggdrasil cluster. 
We acknowledge use of the software package {\tt pycbc}~\cite{pycbc}.
V.DL., G.F. and  A.R. are supported by the Swiss National Science Foundation (SNSF), project {\sl The Non-Gaussian Universe and Cosmological Symmetries}, project number: 200020-178787.
P.P. acknowledges financial support provided under the European Union's H2020 ERC, Starting 
Grant agreement no.~DarkGRA--757480, and under the MIUR PRIN and FARE programmes (GW-NEXT, CUP:~B84I20000100001), and 
support from the Amaldi Research Center funded by the MIUR program `Dipartimento di 
Eccellenza" (CUP:~B81I18001170001).

\appendix

\section{Binary detectability}
\renewcommand{\theequation}{A.\arabic{equation}}
\setcounter{equation}{0}
\label{app:A}

\noindent
A compact quasi-circular binary is characterized by the BHs masses $m_1$ and $m_2$, dimensionless spins $\boldsymbol{\chi}_1$ and $\boldsymbol{\chi}_2$, and by the merger redshift $z$.
We have neglected the role of spins in computing the BH binary detectability in the present work. This is a conservative assumption since spinning binaries typically have larger Signal-to-Noise Ratio~(SNR) than nonspinning binaries with same masses and redshift and, in any case, the large majority of the GWTC-2 events are compatible with zero spin.  
Each individual binary is also characterised by the position and orientation with respect to the detectors. Those are customarily defined in terms of right ascension $\alpha$, declination $\delta$, orbital-plane inclination $\iota$, and polarization angle $\psi$.  
One can thus designate the intrinsic and extrinsic parameters as
 $\theta_\text{\tiny i}=\{m_1,m_2,z\}$ and $\theta_\text{\tiny e} = \{\alpha,\delta,\iota,\psi\}$, respectively.

One has to compute the detectability by averaging over the extrinsic parameters $\theta_\text{\tiny e}$.  For each value of $\theta_\text{\tiny i}$, one can define the detection probability as
\begin{equation}
p_\text{\tiny det}(\theta_\text{\tiny i})= \int  p(\theta_\text{\tiny e}) \, \Theta[\rho(\theta_\text{\tiny i},\theta_\text{\tiny e}) - \rho_\text{\tiny  thr}] \, d \theta_\text{\tiny e}\,,
\label{pdetdefine}
\end{equation}
where $p(\theta_\text{\tiny e})$ is the probability distribution function of $\theta_\text{\tiny e}$,  $\Theta$ indicates the Heaviside step function and $\rho$ is the SNR. 
Notice that, in the case of the GWTC-1 catalog, it was already shown that $p_\text{\tiny det}$ can be computed in the approximate  single-detector semianalytic framework of Refs.~\cite{Finn:1992xs,Finn:1995ah} and a SNR threshold $\rho_\text{\tiny thr}=8$ without encountering significant departures from the large-scale injection campaigns by the LVC for the O1~\cite{Abbott:2016nhf} and O2~\cite{LIGOScientific:2018jsj} runs. We will adopt the same procedure to compute the detectability of binaries also for the O3 run.
 
One can factor out  the dependency of $\theta_\text{\tiny e}$ on the SNR to obtain  $\rho(\theta_\text{\tiny i},\theta_\text{\tiny e})=\omega(\theta_\text{\tiny e}) \rho_\text{\tiny opt}(\theta_\text{\tiny i})$, where $\rho_\text{\tiny opt}$ is the SNR of an ``optimal'' source located overhead the detector with face-on inclination. 
Computing the marginalized distribution $p_\text{\tiny det}(\theta_\text{\tiny i})$ is achieved by evaluating the cumulative distribution function $P(\omega_\text{\tiny thr})=\int_{\omega_\text{\tiny thr}}^1 p(\omega') d\omega'$ at  $\omega_\text{\tiny thr}=\rho_\text{\tiny thr}/\rho_\text{\tiny opt}(\theta_\text{\tiny i})$.

In this work we consider the case of  isotropic sources where $\alpha, \cos\delta, \cos\iota$, and $\psi$ are uniformly distributed. Then, for the case of a single detector approximation, non-precessing sources, and considering only the dominant quadrupole moment, the function $P(\omega_\text{\tiny thr})$ is found as in Ref.~\cite{Dominik:2014yma}.

\begin{figure}[t!]
	\centering
	\includegraphics[width=0.49 \linewidth]{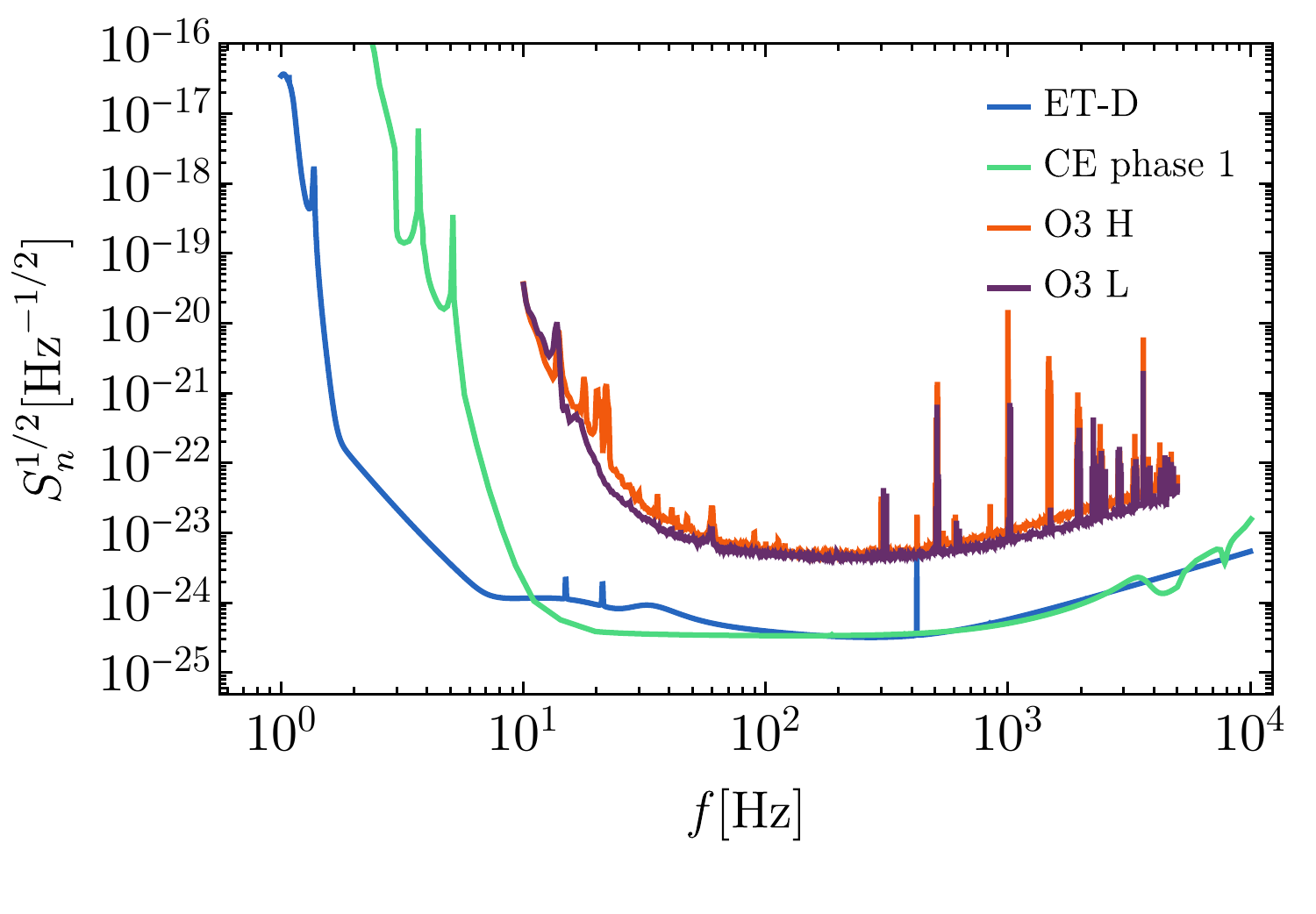}
	\includegraphics[width=0.49 \linewidth]{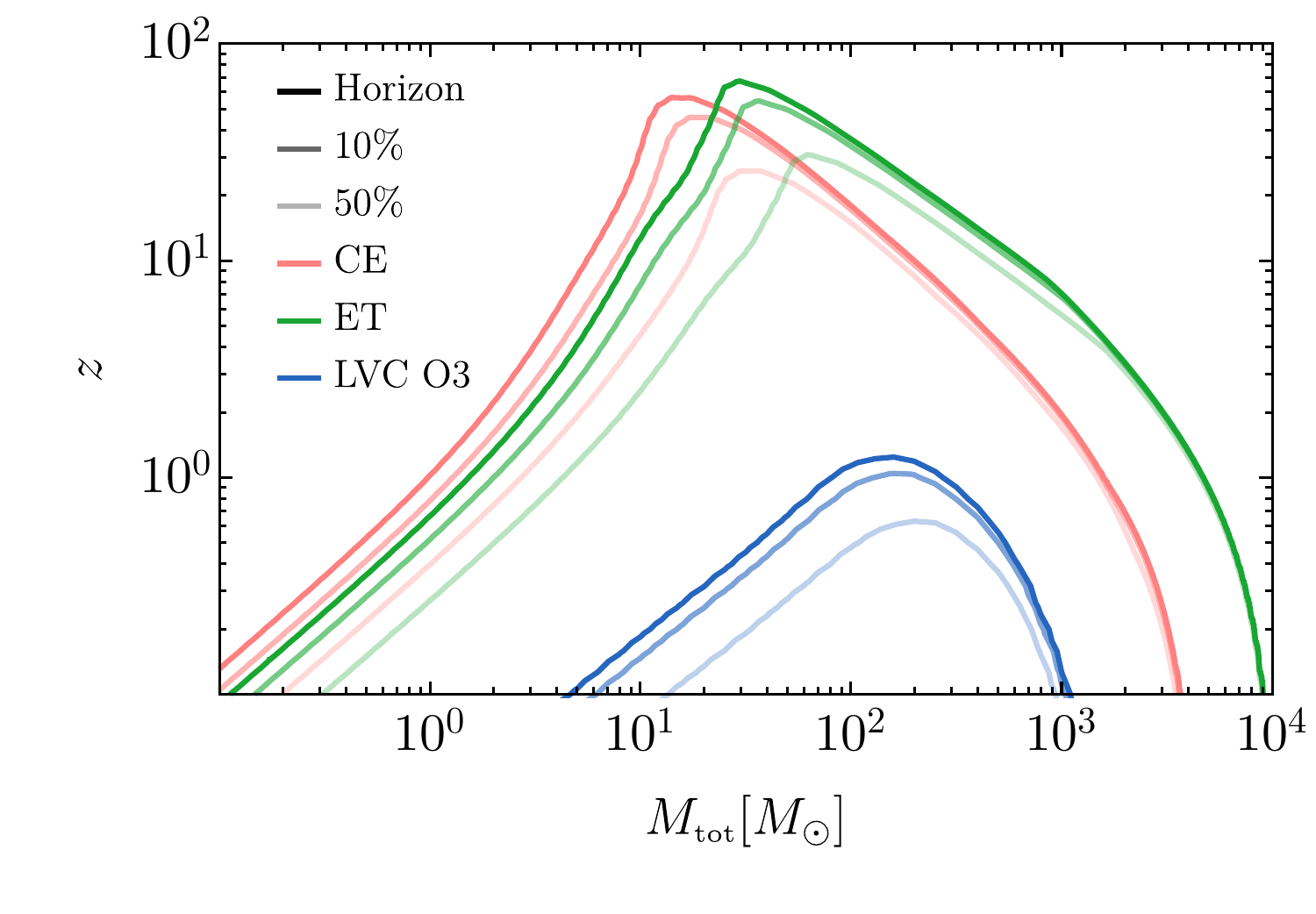}
	\caption{\it \textbf{  Left:} Noise curves.   \textbf{ Right:} Horizon redshift for equal mass binaries.
	Both panels shows the result for the LIGO Hanford (H) and Livingston (L) during the O3 run~\cite{Abbott:2020niy} and both the 3G detectors Cosmic Explorer (CE) during phase 1 from~\cite{ce} and ET at design sensitivity from~\cite{et}. }
	\label{psd}
\end{figure}

The optimal SNR $\rho_\text{\tiny opt}$  of individual GW events for a source with masses $m_1$ and $m_2$ at a  redshift $z$ is given in terms of the GW waveform in fourier space $\tilde h(\nu)$ by 
\begin{equation}
\rho_\text{\tiny opt}^2 (m_1,m_2,z)\equiv \int_0^\infty \frac{4 |\tilde h (\nu)|^2}{S_n (\nu)} {\rm d} \nu,
\end{equation}
where the adopted strain noise $S_n$ for current LIGO/Virgo observation runs and future 3G detectors (such as Einstein Telescope and Cosmic Explorer) are shown in Fig~\ref{psd}. 
In particular, when computing the binary detectabiliy during the  O1-O2 (O3a) observation runs, we adopted the \texttt{aLIGOEarlyHighSensitivityP1200087} (\texttt{aLIGOMidHighSensitivityP1200087}) noise power spectral densities, as implemented in the publicly available repository \texttt{pycbc}~\cite{pycbc}.
Consistently with the computational framework described to compute the binary detectability, we also adopt the non-precessing waveform model IMRPhenomD. Notice finally that the optimal SNR scales like the inverse of the luminosity distance, i.e. $\rho \propto 1/ D_\text{\tiny L}(z)$.

In Fig.~\ref{psd} we also plot the SNR expected at current and future GW experiment for an optimally oriented equal mass binary with total mass $M_\text{\tiny tot}$. 
In the plot we refer to the ``Horizon'' as the maximum distance at which a binary can in principle be observed if optimally oriented with respect to the detectors, while $10\%$,  $50\%$ as the redshift at which those fraction of binaries are observable.
Those values corresponds to values of ${\rm SNR}=\{ 8,10,19 \}$, respectively. 

\section{Accretion onto PBHs in binaries}
\renewcommand{\theequation}{B.\arabic{equation}}
\setcounter{equation}{0}
\label{app:B}

\noindent
PBHs form deep in the radiation dominated early universe, being their mass distribution described by  Eq.~\eqref{psi}.  
Throughout their evolution up to the merger epoch, however, such mass 
function must be properly evolved~\cite{Ricotti:2007jk,Ricotti:2007au,zhang,DeLuca:2020bjf} by taking into account the impact of baryonic accretion.

PBHs in binaries may experience sizeable mass changes if they are heavier than
 $\mathcal{O}(10) M_\odot$. 
 The binary system as a whole attracts baryonic gas from the surrounding background medium depending on  the binary's Bondi-Hoyle mass accretion rate 
\be
 \label{R1bin}
\dot M_\text{\tiny bin} = 4 \pi \lambda m_H n_\text{\tiny gas} v^{-3}_\text{\tiny eff} M^2_\text{\tiny tot}.
\ee
The Bondi-Hoyle rate is computed as a function of the hydrogen mass $m_H$ and mean density $n_\text{\tiny gas}$ while also 
keeping track of the binary effective velocity $v_\text{\tiny eff}$.
The additional factor $\lambda$ is introduced to parametrise corrections to the mass accretion rate coming from the Hubble expansion, the  coupling of the gas to the CMB radiation induced by Compton scattering and, finally, the gas viscosity. The interested reader can find more details in~\cite{Ricotti:2007jk} and references therein. 
The individual mass accretion rates of the PBHs participating in the binary system are modulated by the orbital motion 
and can be parametrised in terms of the mass ratio $q$ as~\cite{DeLuca:2020bjf, DeLuca:2020qqa} 
\begin{align}
\dot m_1 = \dot M_\text{\tiny bin}  \frac{1}{\sqrt{2 (1+q)}}, \qquad \dot m_2 = \dot M_\text{\tiny bin}  \frac{\sqrt{q} }{\sqrt{2 (1+q)}}.
\label{M1M2dotFIN}
\end{align}
We note here that the accretion formula~\eqref{R1bin} is derived in the Newtonian approximation. 
It was recently suggested in Refs.~\cite{Cruz-Osorio:2020dja,Cruz-Osorio:2021qbr}
that there may be an increase of the mass accretion rate due to general-relativistic effects.

It is of crucial importance to account also for the catalysing effect of an additional DM component other than PBHs~\cite{Ricotti:2007au,Adamek:2019gns,Mack:2006gz}.
The most up to date constraints on the DM fraction which is allowed to be in the form of PBHs in the relevant mass range highlight the necessity for a secondary DM component which would tend to cluster around the isolated  PBHs or PBHs binaries.
In order to account for the enhanced gravitational potential well catalysing the baryonic mass accretion we included this effect in the parameter 
$\lambda$~\cite{Ricotti:2007au,DeLuca:2020bjf}. 

The accretion rate is expected to decrease sharply with the onset of structure formation and reionization~\cite{Hasinger:2020ptw,raidalsm,Ali-Haimoud:2017rtz}. We account for this effect by placing an effective cut-off 
in redshift $z_\co$ after which we neglect accretion effects. There exists however large uncertainties in the details of accretion around the reionization epoch as, for example, the effect of PBH induced
X-Ray pre-heating~\cite{Oh:2003pm}, 
details of the structure formation and both local and global feedbacks 
\cite{Ricotti:2007au,Ali-Haimoud:2016mbv} as well as mechanical feedbacks~\cite{Bosch-Ramon:2020pcz}
which force us to consider the cut-off redshift $z_\co$ as an essentially free parameter in the model.

To summarise, the physical effects of mass accretion onto PBHs are expected to be the following:
\begin{itemize}
\item An overall shift of the mass distribution to larger values as well as an enlargement of the high-mass tail ~\cite{DeLuca:2020bjf};
\item A redshift dependent modification of the PBH abundance with respect to the DM proportional to $f_\PBH (z)/f_\PBH(z_\ii) \sim \langle m(z) \rangle/\langle m (z_\ii)\rangle$~\cite{paper2};
\item PBH binaries tend to become more symmetrical (i.e, $q$ closer to unity), depending on the strength of accretion, as the secondary PBH in the binary is expected to consistently inherit a larger relative accretion rate;
\item mass accretion may lead to an efficient spin-up of both PBHs in the binary provided enough angular momentum is transferred to the PBH~\cite{bv,Shakura:1972te,NovikovThorne,Bardeen:1972fi,Brito:2014wla,volo,thorne,Gammie:2003qi} (as it is the case for a disk-like accretion geometry). For more details on the spin accretion in the PBH model we refer to Refs.~\cite{DeLuca:2020bjf,DeLuca:2020qqa}.
\end{itemize}

\end{document}